\documentclass[journal,onecolumn]{IEEEtran}
\pdfoutput=1

\usepackage{graphicx}
\usepackage{subfig}
\usepackage{amssymb, amsmath,amsthm, amsfonts}
\usepackage{ifthen}
\usepackage{tikz}
\usepackage{enumerate}
\usepackage{verbatim}
\usepackage{pifont}
\usepackage[justification=centering]{caption}
\usepackage{bm}  

\DeclareSymbolFont{bbold}{U}{bbold}{m}{n}
\DeclareSymbolFontAlphabet{\mathbbold}{bbold}

\newtheorem{theorem}{Theorem}

\newtheorem{lemma}[theorem]{Lemma}

\newtheorem{remark}[theorem]{Remark}
\newtheorem{definition}[theorem]{Definition}
\newtheorem{example}{Example}
\newtheorem{claim}[theorem]{Claim}

\newcommand{\be}{\begin{equation}}
\newcommand{\ee}{\end{equation}}
\newcommand{\ben}{\begin{equation*}}
\newcommand{\een}{\end{equation*}}
\newcommand{\ba}{\begin{eqnarray}}
\newcommand{\ea}{\end{eqnarray}}

\newcommand{\indicator}[1]{\mathbbold{1}_{\{ {#1} \} }}

\newcommand{\ssuma}[2]{\underset{j\in S_1[#2]}{\sum }}
\newcommand{\ssumb}[2]{\underset{j \in S_2[#1]}{\sum }}

\newcommand{\ssum}[2]{\underset{j=#1}{\overset{#2}{\sum\!\!\!\!\!\!\!\!\!\circledS\,\,\,\,}}}

\def\h2{\tilde h}

\def\hm1{\hat h_{-1}}

\def\mi{{h_i(0)}}
\def\mione{h_{1}(0)}
\def\mitwo{h_{2}(0)}
\def\mia{h_{i}(0)}
\def\mias{h_{i}^2(0)}
\def\sm{S_m}
\def\boh{\mathbf{h}}
\def\boH{\mathbf{H}}
\def\boP{\mathbf{P}}
\def\boR{\mathbf{R}}
\def\bow{\mathbf{w}}
\def\bpsi{\bm{\psi}}
\def\eE{\mathbb E}

\DeclareMathOperator{\argmax}{argmax}

\title{Distributed Rate Adaptation and Power Control in Fading Multiple Access Channels}
\author{
  	Sreejith~Sreekumar\authorrefmark{1}, 
	Bikash~K~Dey\authorrefmark{1},
	Sibi~Raj~B~Pillai\authorrefmark{1} 

\thanks{This paper was presented in part at the International Symposium on Information
Theory, ISIT 2013, Istanbul and also at the Information Theory Workshop, ITW 2014, Tasmania.}

\thanks{\authorrefmark{1}The authors are with the Department of Electrical Engineering at
IIT Bombay, Mumbai, INDIA-400076. Email:\{sreejiths, bikash, bsraj\}@ee.iitb.ac.in}

}

\begin{document}

\maketitle 

\usetikzlibrary{arrows}

\begin{abstract}
Traditionally, the capacity region of a coherent fading  multiple access channel (MAC) is 
analyzed in two popular contexts. In the first, a centralized system with  
full channel state information at the transmitters (CSIT) is assumed, and the communication parameters like
transmit power and data-rate are jointly chosen  for every fading vector
realization. On the other hand, in fast-fading links with distributed CSIT, the lack
of full CSI is compensated by performing ergodic averaging over sufficiently many
channel realizations. Notice that the distributed CSI may necessitate decentralized power-control
for optimal data-transfer.  Apart from these two models, the case of slow-fading links 
and distributed CSIT,  though relevant to many systems, has received much less attention.

In this paper, a block-fading AWGN MAC 
with full CSI at the receiver and distributed CSI at the transmitters is considered.
The links undergo independent fading, but otherwise  have arbitrary fading distributions.
The channel statistics and respective long-term average transmit powers are known to all
parties. 
We first consider the case where each encoder has knowledge only of its own 
link quality, and not of others. For this model, we compute the adaptive capacity 
region, i.e. the collection of average rate-tuples under block-wise coding/decoding 
such that the rate-tuple for every fading realization is inside
the instantaneous MAC capacity region. 
The key step in our solution is an optimal rate allocation function for any
given set of distributed power control laws at the transmitters. This also allows 
us to characterize the  optimal power control for a wide class of fading models. 
Further extensions are also proposed to account for more  general CSI availability at the transmitters.

\end{abstract}

\section{Introduction}\label{sec:intro}
The multiple access channel (MAC) is a fundamental model for many 
multiple-transmitter single-receiver systems, such as the
up-link of a cellular network. 
It is well known that the achievable data-rates over 
a fading MAC system depends on the availability of channel state 
information (CSI). While it is reasonable to assume that the receiver has
access to full CSI, the availability of the CSI at the transmitters (CSIT) 
depends on 
factors like the coherence-time, admissible feedback overhead  etc. 
In this paper, we consider a slow fading MAC with distributed CSI at the 
transmitters and full receiver CSI. 
We call this a distributed CSI MAC, where each encoder has
some level of local CSI available.

There has been significant work on fading MAC channels under different CSI
assumptions at the encoders.  
In fast fading channels, coding over a large block spanning many
independent fading states is common, and it brings the average behaviour
of the channel into play in the same coding block. The resulting capacity
region is called the ergodic capacity region. 
The ergodic capacity region for a fading AWGN MAC has been characterized under
perfect CSI at the transmitters and the receiver~\cite{TseHanly98, TseViswanath05}. 
A distributed CSI model where each encoder is aware only of its own link quality 
is considered in \cite{ShamaiTelatar99}, where the ergodic sum-capacity is analyzed.
Under more generalized CSI availability at the transmitters, 
\cite{DasNarayan02} characterized the ergodic capacity region as an
optimization problem over suitable power-control laws. However, explicit
solutions for the optimal power-control are difficult to obtain, and 
good thumb rules for distributed power control are usually employed~\cite{ShamaiTelatar99}. 
MACs with fast-fading can also be analyzed  using the framework of 
channels with state. Models of discrete memoryless MAC with state have got
significant attention under various assumptions on CSI availability,
such as causal/non-causal CSIT \cite{KeStMe08}, \cite{LaSt13},
asymmetric CSIT \cite{CemalSteinberg05}, \cite{Jafar06},  
asymmetric CSI at the transmitters and no CSI at the receiver~\cite{ZaPiSh13}
etc. 
Notice that a ergodic utility is more suitable in a fast-fading model,
where sufficient channel variations are available in the coding block.
For slow fading models with full CSIT, the results of \cite{TseHanly98} 
still apply, and the capacity region is known. A remaining question of
 interest  is on slow-fading models with distributed CSI. 

Consider a block fading AWGN MAC, where the fading states remain fixed for a large
block length (coherence time), and change in an i.i.d. manner from block
to block, a widely used assumption \cite{BiPrSh98}. 
Unlike in the ergodic framework, coding is allowed only within a 
single block or coherence time, which is assumed to be large enough.
Such within block coding models appear in several
practical slow fading contexts~\cite{HaTs98}, \cite{CaTaBi99}.
In addition, the transmitters may have varying levels of CSI availability, leading
to a distributed CSI MAC.
There are two possible modes of operation under distributed CSIT, 
as described below.

\noindent \textbf{(i) Safe Mode:} In this mode, henceforth also called the `{\it outage-free}' mode,
the transmitters attempt to play it safe in each block, by choosing rates
and powers
such that the data can be decoded at the receiver. The challenge
is to choose the rates and powers blockwise based on the distributed CSI, 
while ensuring correct decoding  with high enough probability in
each block.
Such a MAC model was  introduced in  \cite{HwMaGaCi07}, \cite{GamalKim11}, where
the case of distributed state information at the respective encoders was considered.
Note that, in contrast, the ergodic setup requires
the error probability to be low when averaged over a large number of fading 
realizations.
We assume a sufficiently large block length (coherence time), and require that
the average error probability  decays exponentially to zero in blocklength for every 
fading vector realization. In other words, the rate-tuple in each block should be within 
the instantiated  MAC capacity region, which is determined by 
the fading realizations and the chosen transmit powers in that block. We will
say that the system remains `outage-free' in each block.
The long-term average (over blocks) rate-tuples achievable under this model is known
as the \emph{adaptive capacity region}.

\noindent \textbf{(ii) ARQ mode:} Another option in the distributed CSI setting
is to adopt a more aggressive
rate-choice which allows the effective rate-tuple to be outside the instantaneous
capacity region for some combinations of the channel states. We call such events as outage (to be
defined more precisely later) and these events may result in a  high probability
of error in the respective blocks. The lost data can either be re-transmitted
in an ARQ based system with feedback or can be recovered  using an
inter-block outer erasure code. It may be noted that the inter-block outer 
erasure code violates the basic framework of within-block coding, and is a special
form of coding across fading states. In either case, the achieved rate is
calculated by simply discounting the lost data in the outage events.
The capacity region under this setup will be inside the ergodic
capacity region, but may be bigger~\cite{DPD11} than the outage-free capacity 
region for the `safe-mode'. Alternate approaches based on broadcasting
to mitigate the lack of CSIT also exists, see \cite{MiFrTs12} for a recent
account.

For both safe mode as well as ARQ mode, there are two time-scales of interest. In the
terminology of \cite{BiPrSh98}, a `short-term' or per-block average power
constraint dictates the choice of codebooks used in a block. The transmitter
may have some freedom in adapting the short-term constraint based on the available
CSI, however the adaptations should respect a long-term average (over blocks)
power constraint imposed by 
physical considerations. We will use the same nomenclature here, see also
\cite{CaTaBi99}, \cite{HaTs98} for the origin and physical significance of these terms.
Similarly, the rate-adaption schemes may change the transmission-rates from block to block,
and our utilities capture the long-term average rates.

As in~\cite{GamalKim11,HwMaGaCi07,DPD11,IPD12}, this paper focuses on outage-free (\emph{safe-mode})
operations over block fading MAC under distributed CSI. 
For most parts of this paper, we consider a fading MAC where each transmitter is aware only of its own link quality,
we call this the \emph{individual CSI} MAC. 
This type of distributed CSI at the transmitters
is  practical in various setups \cite{GamalKim11}(page~590--593),
for instance, when the channels are estimated
by the transmitters during the downlink broadcast phase of a time-division
duplex (TDD) mode operation. Notice that \cite{ShamaiTelatar99} 
considered the
same individual CSI model, however the ergodic sum-capacity under fast-fading
was the utility of interest there. As we mentioned earlier, the adaptive capacity
is the region of interest in the safe mode.
The lack of global CSI calls for novel access schemes to maximize data-transfer. 
These schemes should facilitate 
each transmitter to exploit its channel knowledge  in increasing the individual
data-rate, at the same time not resulting in an outage for
any possible fading state of the other links.
Communication techniques should account for the tension between these two 
competing requirements.
%

The early works~\cite{GamalKim11, HwMaGaCi07} 
gave a formulation of the adaptive capacity region as an optimization whose numerical 
evaluation is only tractable for a small number of discrete fading 
states. The terminology \textit{adaptive capacity region} was introduced
in \cite{GamalKim11}.
While the adaptive capacity region as such is defined for fixed
transmit powers at the respective encoders, more flexibility can be made
available by adapting the transmit powers, the resulting 
utility is known as {\em power-controlled adaptive 
capacity region} \cite{GamalKim11}. We will normally use the former terminology for
both the utilities, either the reference will be clear from the context, or we may
append the word \emph{power} to signify power control.
Recently, the adaptive sum-capacity under identical
fading statistics across users were presented in~\cite{DPD11}, \cite{IPD12}, 
where the optimal power-allocation was shown to have a water-filling 
form.
It was also shown in~\cite{DPD11} that the sum-capacity can be achieved
by rate-splitting and a successive cancellation decoder of lower complexity.
The main contribution of the current paper is in characterizing the
complete adaptive capacity-region of an individual CSI MAC, valid for
 arbitrary fading statistics and power constraints. Extensions to
other local CSI models are also proposed.

%

\begin{figure}[htbp]
\centering \includegraphics[scale=0.9]{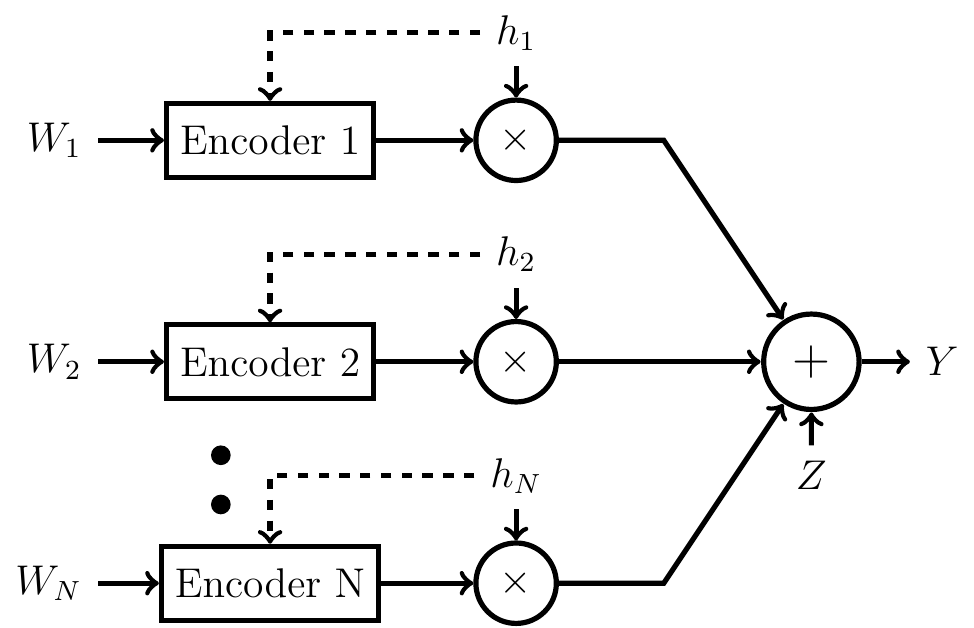}
\caption{ Gaussian fading MAC with Individual CSI at transmitters and full CSI at receiver \label{3indcsi}}
\end{figure}


\subsection{Contribution and organization of this paper}
This work  primarily addresses the power controlled adaptive capacity 
region for an \emph{individual CSI MAC} under arbitrary fading distributions,
independent across links. Section~\ref{sec:model} presents the system model together
with some definitions and notations.
We summarize our contribution below with respect to earlier related
works.
\begin{itemize}
\item For a given set of power control laws at the transmitters, 
we present an almost closed form solution in Section~\ref{sec:sum_capacity} to 
the \emph{adaptive sum-capacity} for the distributed
 CSI MAC with arbitrary fading distributions, which are independent across links. 
Presented for both discrete
and continuous fading states, these are easily computable for any set of
fading distributions. In contrast, earlier 
works like \cite{GamalKim11}, \cite{HwMaGaCi07} focused more on a single
letter characterization for the discrete memoryless case. Evaluating
these formulas for the Gaussian case  resulted in 
unsolved optimization problems in terms of power control and rate-adaptation
functions.  Notice that simple numerical solutions for such problems  
can  only handle channels with very few states and a small number of users.
The work in \cite{DPD11,IPD12} provided the solution for some special
cases. The approach there  critically depends on the assumption of 
identical channel statistics across users, a limitation which is 
circumvented in this work using a novel rate-adaptation technique.

For accessibility, we will describe the results for the  discrete fading states 
in detail first, given in 
Sec.~\ref{subsec:disc:two} (for two users) and Sec.~\ref{subsec:dis:mult} 
(for arbitrary number of users). These are then generalized
to cover the corresponding continuous valued fading states, in Sec.~\ref{subsec:cont} (for two users) 
and in Sec.~\ref{subsec:cont:mult} (for arbitrary number of users).
The two users case is presented first for both types of fading states
since this clearly illustrates the underlying ideas. The
generalization to multiple users also require some new techniques in the
proofs.
While generalized formulae
encompassing both discrete and continuous valued fading states are possible, 
it will make the presentation a bit awkward. Furthermore,
the discrete case carries considerable insight into the solutions, in addition to the chronological order
in which the results were obtained. 
\item Section~\ref{sec:adap:capa:reg} generalizes the results of 
Section~\ref{sec:sum_capacity} to find the maximum weighted sum-capacity
for any weight vector, thus allowing the computation of the whole adaptive 
capacity region by taking different weight vectors.
\item In section~\ref{sec:PCReg}, we present the power-controlled adaptive 
capacity region as a convex optimization problem under linear constraints
for discrete (finite number
of) fading states. The optimization problem is shown to
be tractable due to a crucial monotonicity property for an optimal power
allocation. It is shown that there is an optimal power allocation function
for which the received power is a monotonically non-decreasing function of the
fading magnitude. This allows the weighted sum-rate to be expressed as
a fixed function of the power allocation, and this leads to an optimization
problem where the number of variables (power values) is same as the number
of states. 
\item In section~\ref{sec:extra:csi}, we extend the results to 
a CSI model where each user knows
some partial information about the other users' fading states. 
The proposed techniques also easily extend to the case of
arbitrary CSIT models provided that the knowledge of its own state at a user
is at least as good as other users' knowledge of the same. In other words,
each user is aware of what others know about its fading state.
It is worth pointing out that the two user asymmetric CSI MAC model of \cite{CemalSteinberg05}
is an extreme case of the CSI availability that we consider, 
where one user has full CSIT, and the other knows only
its own link. Single letter characterizations for the ergodic region
of asymmetric CSI models are available \cite{CemalSteinberg05}, see 
\cite{Jafar06} for generalizations.
In contrast, we consider adaptive coding under more generalized versions of CSI availability. 
Nevertheless, the techniques that we propose in Section~\ref{sec:extra:csi}
also allow the numerical evaluation of 
the capacity region for specific cases like the asymmetric CSI MAC of \cite{CemalSteinberg05},
a result of independent interest. This connection is not further explored in the current paper.
\end{itemize}
Finally, Section~\ref{sec:conclusion} concludes the
paper with suggestions for some future work.

\section{System Model}\label{sec:model}
Consider a system where 
$N$ transmitters have independent data-streams to be sent to a common receiver. 
We use the subscript $i\in\{1,2,\cdots, N \}$  to represent variables associated with user~$i$. 
The channel is modeled as a block fading MAC where the received symbol
is given by
\begin{align}
Y = \sum_{i=1}^N H_i X_i + Z,
\end{align}
where $X_i \in \mathbb R$ is the symbol transmitted by user $i$, $H_i \in
\mathcal H_i \subseteq \mathbb R$  is the fading state of the channel from
user $i$ to the receiver, and $Z \sim \mathcal N (0,1)$ is a real additive white
Gaussian noise (the variance is assumed to be $1$ without loss of generality) independent of $\{X_i| 1\leq i\leq N\}$ and $\{H_i| 1\leq i\leq N\}$. 
The fading coefficients $H_i,~1 \leq i\leq N$ are assumed to be independent of each other.
The fading vector $\boH=(H_1, H_2, \cdots, H_N)$ remains constant within a sufficiently 
large block of fixed size and varies independently across 
blocks. We assume that the fading statistics as well as the respective long-term average 
power constraints  are known to all parties. 

The transmitters have the freedom to adapt their rates and power according to the 
available local knowledge of the fading vector. However the choice of rates should 
ensure that the decoding error probability exponentially decays with blocklength
for every realization of the fading vector. This is different from having an 
arbitrarily small error  probability in the Shannon sense,  which may need infinite block-lengths, 
see \cite{GamalKim11}(page~$587$). In particular, our target is an acceptably small
error probability permitted by the large blocklength, while averaged over the
uniform choice of messages in that block. 

We assume a CSI model where the $i$-th user has an estimate
$\hat{H}_j^{(i)} := g_{ij}(H_j)$ of $H_j$, where $g_{ij}$ is a function. 
So the CSI available at the
$i$-th user is $\hat{\boH}^{(i)}:=(\hat{H}_1^{(i)}, \hat{H}_2^{(i)}, \cdots, \hat{H}_N^{(i)}\}$.
Note that the estimates $\hat{H}_j^{(i)}$ are deterministic functions of $H_j$,
and there is no random noise in the estimate.
For most parts of this paper (Sec.~\ref{sec:sum_capacity} till Sec.~\ref{sec:PCReg}) we assume
that the $i$-th transmitter knows only its own channel state $H_i$ before
transmitting in that block. That is, $\hat{H}_i^{(i)} = H_i$ $\forall i$,
and $\hat{H}_j^{(i)} =\emptyset, \forall j\neq i$. In Sec.~\ref{sec:extra:csi}, we will 
relax our assumptions and equip
user~$i$ with some additional partial information about the other channel states
$H_j; j\neq i$, i.e., $g_{ij}$ is not a constant function for all $j\neq i$.
By an abuse of notation, we will denote the image of $g_{ij}$ by $\hat{\mathcal H}_{j}^{(i)}$, and so $\hat{H}_j^{(i)} \in \hat{\mathcal H}_{j}^{(i)}$.

A few more comments on notation are in order.
We will denote vectors by bold-face, i.e. $\textbf u$ represent a vector with $u_i$ at position~$i$, where $u_i$
can be either a scalar or a function. The overbar symbol usually denotes an average quantity.
Also, in case of multiple subscripts, we may write $h_{i,j+k}$ as $h_{i(j+k)}$ for clarity.
%
%

The following definitions are given for the general CSI model  described
above, though we will mostly consider the special case of an individual  CSI MAC.
\begin{definition}
 A power rate strategy is a collection of mappings $(P_i,R_i):
\hat{\mathcal H}_{1}^{(i)} \times \hat{\mathcal H}_{2}^{(i)} \times \cdots \times \hat{\mathcal H}_{N}^{(i)} \rightarrow \mathbb{R}^+ \times \mathbb{R}^+ \times
\cdots \mathbb{R}^+ ,\, 1\leq i \leq N$.
\end{definition}
Thus, in the  global fading-state $\boH$, the $i^{th}$ user employs 
a codebook of rate $R_i(\hat{\boH}^{(i)})$ and power $P_i(\hat{\boH}^{(i)})$.
Let $ C_{MAC}(\boh, \boP)$ denote the capacity region of a 
Gaussian multiple-access  channel  with a fixed fading vector $\boh$  
and average power-constraint $P_i$ for the user $i$, $1\leq i \leq N$. 
It is well known \cite{CovTho91,GamalKim11} that  $C_{MAC}(\boh, \boP)$ is the collection of all 
rate-tuples of the form $\boR = (R_1,R_2,\cdots, R_N)$ such that
\begin{align}\label{eq:mac:pent}
\forall S \in \{1,\cdots, N\}, \sum_{i\in S}R_i \leq \frac 12 \log\left(1+\sum_{i\in S}h_i^2 P_i\right).
\end{align}

\begin{definition}
A power-rate strategy is called feasible if it satisfies the
average power constraints of the users, i.e.  for $1\leq i \leq N$,
$\eE_{\boH}\left(P_i\left({\hat{\boH}}^{(i)}\right)\right) \leq P_i^{avg}$,
where $P_i^{avg}$ is the long-term average power constraint of user~$i$
and $\mathbb E(\cdot)$ denotes the expectation operator.
\end{definition}

\begin{definition} \label{def:out:free}
 A power-rate strategy $(P_1(\cdot),R_1(\cdot),\cdots, P_N(\cdot),R_N(\cdot))$ is termed as outage free if \\
$$
\forall \boh \in \{\mathcal{H}_1 \times, \cdots,\times \mathcal H_N\}, 
\left( R_1(\hat{\boH}^{(1)}),\cdots, R_N(\hat{\boH}^{(N)}) \right) \in 
	C_{MAC}(\boh,P_1(\hat{\boH}^{(1)}),\cdots, P_N(\hat{\boH}^{(N)})).
$$   
\end{definition}
Such an outage-free power-rate strategy ensures that in each block,
the rate-tuple chosen distributedly by the users is inside the polymatroid
capacity region given in \eqref{eq:mac:pent},  under the distributed choice of powers
$P_1(\hat{\boH}^{(1)}),\cdots, P_N(\hat{\boH}^{(N)})$. Thus, Gaussian codebooks
at these rates can achieve a decoding error probability exponentially
decaying to zero with block-length. The long-term average achieved rate of user~$i$ for a 
given power-rate allocation strategy is given by
\begin{align*}
\bar R_i := \mathbb E \left[ R_i(\hat{\boH}^{(i)}) \right]
\end{align*}
where the expectation is over $\boH$. The average rate-tuple achieved
by a power-rate strategy is then $\bar {\boR} = ({\bar R}_1, \bar R_2,
\cdots, \bar R_N)$.
Let $\Theta_{MAC}(\boP^{avg})$
denote the collection of all \textbf{feasible} power-rate  strategies which are
outage-free.

\begin{definition}
A rate vector $\bar{\boR}$ is said to be an {\it achievable} rate-tuple
under power-rate adaptation if 
there exists a feasible outage-free power-rate allocation strategy for which
the expected rate-tuple is $\bar{\boR}$.
The {\it power-adaptive capacity region} is defined as the closure of the
set of achievable expected rate-tuples under power-rate adaptation.
\end{definition}

The power-adaptive capacity region can be evaluated by computing
the power-controlled adaptive weighted sum-capacity for every non-negative
weight vector as defined below.

\begin{definition}\label{def:w:sum:pow}
 The power-controlled adaptive weighted sum-capacity
$C^{pc}_{sum}(\bow, \bpsi)$, for a non-negative vector  $\bow = (w_1, \cdots, w_N)$
is defined as
$$
C^{pc}_{sum}(\bow, \bpsi)= \max \sum_{i=1}^N \mathbb{E}  \left[ w_i R_i(\hat{\boH}^{(i)}) \right]
$$
where the maximization is over all feasible outage-free power-rate strategies in $\Theta_{MAC}(\boP^{avg})$.
\end{definition}

In some settings,
the adaptation is  limited to the transmit rates in each block, and
the power-control law is specified in advance.
The corresponding expected rate-region is known as the 
\emph{adaptive capacity region}. 
Such schemes are of interest in situations
where good/practical power control laws are already specified based on 
heuristics
or other engineering considerations \cite{ShamaiTelatar99}. 
In several other
systems, a regulatory transmit spectral cap may force the power-control to take
particularly simple forms, for example, a constant power. 
Rate-adaptation is the only freedom available in such situations~\cite{SibiHanly10}.
Notice that in the individual CSIT MAC, a pre-specified power allocation $P_i(H_i)$
is equivalent to no power adaptation, as its effect can be absorbed in
the fading coefficients by considering the new fading state to be $\sqrt{P_i(H_i)}H_i$ 
(with an appropriate distribution on the new fading states).

Though our general interest is to find the power-adaptive capacity region,
we will first develop techniques for the case of constant power 
allocation (or no power control).
For a given set of power-control laws across users, let $\vartheta_{MAC}(\boP^{avg})$
denote the collection of all \textbf{feasible} rate-adaptation strategies which are
\textbf{outage free}. Similar to Definition~\ref{def:w:sum:pow}, the adaptive capacity region 
can be characterized by an equivalent weighted sum-rate maximization, defined below. Let us consider
a fixed transmit power vector $\boP^{avg}$.
\begin{definition}\label{def:w:sum}
The weighted adaptive sum-capacity $C_{sum}(\bow, \bpsi)$, for a 
non-negative vector $\bow = (w_1, \cdots, w_N)$, is defined as
$$ 
C_{sum}(\bow,\bpsi)= \max \sum_{i=1}^N \mathbb{E}  \left[ w_i R_i(\hat{\boH}^{(i)}) \right]
$$
where the maximization is over all feasible outage-free rate strategies in 
$\vartheta_{MAC}(\boP^{avg})$.
\end{definition}
When all the weights $w_i$ are identically one, the sum-throughput is known as
the \emph{adaptive sum-capacity}. This case is of special interest, 
and all our expositions will start with the sum-capacity, and then extended 
to the weighted sum-capacity.



\begin{remark}
While the weighted sum-rates can be used to characterize the entire capacity region, sometimes
a convex hull operation become necessary. However, in the cases that we consider further, the
utilities take the form of an `expectation of logarithm function', and the convex-hull becomes superfluous.  
\end{remark}
%

Note that for the individual CSI model we consider in most of the paper
(from Sec.~\ref{sec:sum_capacity} to Sec.~\ref{sec:adap:capa:reg}), the power allocation functions $P_i(\cdot)$ and
the rate allocation functions $R_i(\cdot)$ are simply functions of
$H_i$.

%
\begin{remark}
Since each transmitter is aware of its link CSI, only fading magnitudes are 
important in the computation of the rates. Thus without loss of generality,
we assume positive valued fading coefficients for the rest of the paper. 
\end{remark} 

The following definition will be very useful for our technical results.
\begin{definition} \label{def:inv:cdf}
The inverse CDF function for user~$i$ is
\begin{align} \label{eq:cdf:inv}
h_i(x) = \psi^{-1}(x) := 
	\begin{cases} \sup\{h |\, \psi_i(h) < x \} \text{ for } 0 < x \leq 1 \\
		0 \textrm{ when } x = 0.
	\end{cases}
\end{align}
\end{definition}
Using this  definition,
we will slightly abuse the notation and express the long term average rate for 
user~$i$ in  an individual CSI MAC as
\begin{align} \label{eq:rate:cdf}
\mathbb E\left[R_i(H_i)\right] =  \int_{0}^1 R_i(h_i(x)) dx, \, 1\leq i \leq N.
\end{align}
In writing the integral, we have implicitly assumed well-behaved fading
distributions, which can be discrete, continuous-valued or mixed.
We now state a simple lemma which finds multiple applications in this paper.
\begin{lemma} \label{lem:log:concave}
Let $(u_1,u_2$ and $(v_1, v_2)$ be two non-negative vectors with $u_1+ u_2 = v_1 + v_2$.
If $u_1 \leq v_1 \leq u_2$ and $u_1 \leq v_2 \leq u_2$, then
$$
\log(1 + u_1) + \log (1 + u_2) \leq \log(1 + v_1) + \log(1+v_2).
$$
\end{lemma}
The lemma follows by the concavity of the logarithm function. The 
stage is now set for presenting our results, and we will start with the
adaptive sum-capacity of an individual CSI MAC in the next section.

\section{Adaptive Sum Capacity Without Power Control}
\label{sec:sum_capacity}

In this section, we consider an individual CSI MAC, where the transmitters
adapt their rates based on the knowledge of their own fading coefficients 
in a distributed  manner. We will start with a model where user~$i$ has a fixed transmit
power of $P_i$. This corresponds to a short-term, per-block, average power
constraint of $P_i$ in every block. This model is also 
considered in \cite{GamalKim11}, where the optimal rate-allocation
is unsolved. 
The significance and applications of blockwise 
short-term  average power constraints in fading models are
detailed in \cite{BiPrSh98}, see also \cite{CaTaBi99}. Furthermore, employing fixed power 
constraints are common in models where there is a spectral cap on the 
transmissions~\cite{SibiHanly10}.
Apart from the significance of the model, 
the solution of the adaptive sum-capacity  problem for fixed powers  
illustrates our key techniques, which will later 
prove useful in computing  the full capacity region as well as the optimal 
power allocation functions. Handling discrete and continuous-valued fading 
distributions need somewhat different treatments. We will first present the 
discrete case,  generalizations to arbitrary distributions are presented  
in Sections~\ref{subsec:cont} and~\ref{subsec:cont:mult}.  
  
For simplicity of exposition, we will first  consider
two user MACs  and later  generalize to the  $N-$users case.
The generalizations require somewhat more involved proofs, however they follow
the same two user principles. 


%

\subsection{DISCRETE FADING STATES: TWO USERS}
\label{subsec:disc:two}

In this section, we develop an inductive algorithm to perform
the  optimal rate allocation for discrete fading states. 
%
Let us consider a two-user fading MAC with 
fading CDFs  $\psi_1(h)$ and $\psi_2(h)$. 
We first consider an example MAC with two states for each link to illustrate 
the idea behind the optimal rate allocation. 

\begin{figure}[htbp]
\centering \includegraphics[scale=0.9]{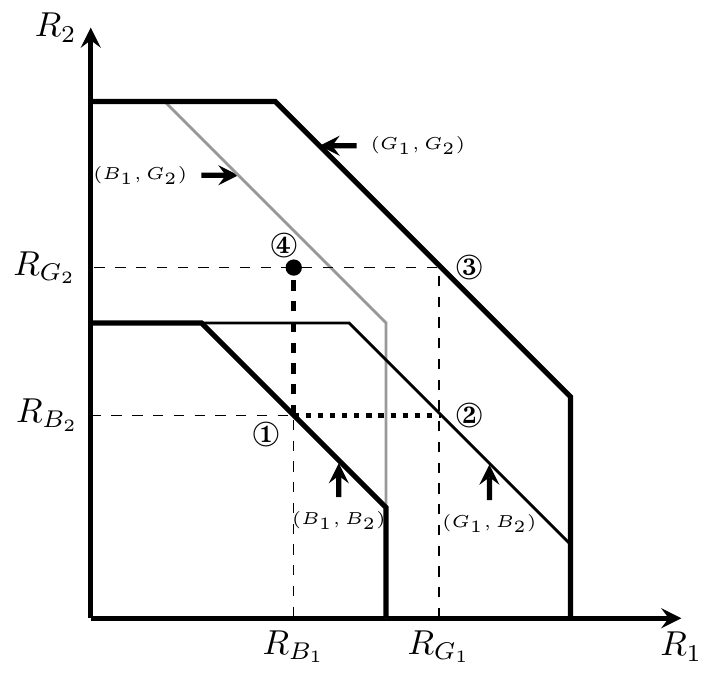}
\caption{Illustrating the rate-choice for a 2 state MAC}
\label{fig:macpent}
\end{figure}

\begin{example}
Let us consider a 2-user MAC, with each link having two states.
The weaker of the states  is
referred to as the bad ($B$) state and the stronger state is referred to as
the good ($G$) state. For link $i$, these are denoted by respectively $B_i$ and $G_i$.
Fig.~\ref{fig:macpent} shows the MAC capacity regions for
each pair of states of the links. 
For example, the inner pentagon is the capacity region for the 
state-pair $(B_1, B_2)$, and outer pentagon is the capacity region for $(G_1, G_2)$.
Our rate-allocation (in Theorem~\ref{th:discrete}) first chooses any point  on the
dominant face of the pentagon for the $(B_1,B_2)$ state-pair and assigns the
respective co-ordinate values to the rates $R_1(B_1), R_2(B_2)$ for the Bad 
state-pair. This point is marked as \ding{172}. 
Suppose that $G_1$ has a higher probability than $G_2$.
Then, we can prioritize the rate $R_1(G_1)$ over $R_2(G_2)$.
Suppose the horizontal line through the point \ding{172} intersects the
pentagon for the $(G_1,B_2)$ state on the dominant face at point \ding{173}.
The horizontal coordinate of this point is assigned as the rate $R_1(G_1)$
for the state $G_1$ of user~$1$. Note that this is the maximum $R_1(G_1)$
(given $R_2(B_2)$) that does not cause outage at the state pair $(G_1,B_2)$.
Now suppose the vertical line through \ding{173} intersects the pentagon for the $(G_1,G_2)$ 
state at point \ding{174} on its dominant face. 
The vertical coordinate of this point determines the rate $R_2(G_2)$ of user~$2$ for the state $G_2$.
The allocation ensures (as will be shown in Lemma~\ref{lem:discrete}) that the operating rate-pair \ding{175} for the
state-pair $(B_1,G_2)$ is also inside the corresponding capacity region, as depicted in Fig.~\ref{fig:macpent}.
\end{example}

%
%

%

\def\hao{g_{1a}} \def\hbo{g_{1b}} \def\hco{g_{1c}}
\def\hatt{g_{2a}} \def\hbt{g_{2b}} \def\hct{g_{2c}}

\begin{figure}
\vspace*{-0.5cm}
\centering\includegraphics{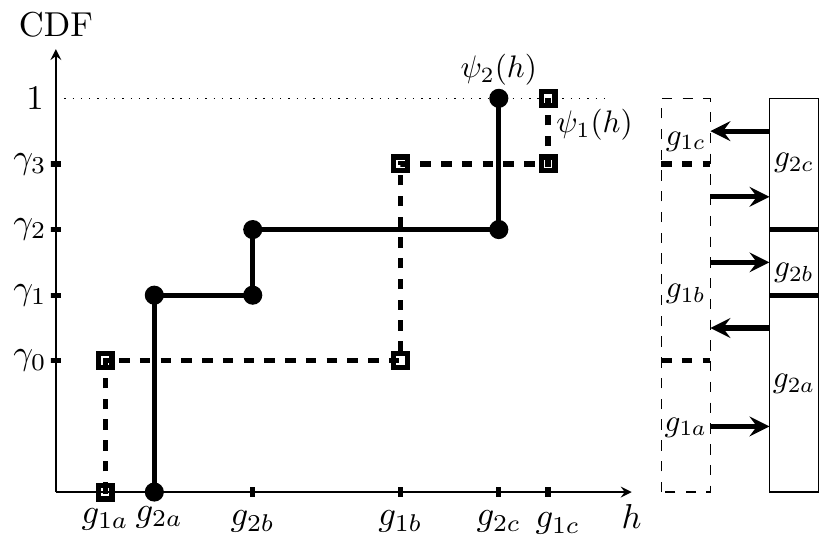}
\caption{Illustrating the rate-assignment for $H_1 \in \{\hao,\hbo,\hco\}$ and $H_2 \in 
	\{\hatt,\hbt,\hct\}$. \label{fig:rate:assign}}
\end{figure}

Now we discuss the rate-allocation for arbitrary discrete states. 
Let user 1 have $k_1$ channel states
with probabilities $p_i; 0\leq i \leq k_1-1$ and let user 2 have $k_2$ 
channel states with probabilities $q_i; 0\leq i \leq k_2-1$. Let us denote
the CDF values of the channels as
\begin{align}
 \alpha_i &= \sum_{j=0}^i p_j\,,\,\, 0 \leq i \leq k_1-1, \label{eq:def:alpha} \\
\beta_i &= \sum_{j=0}^i q_j \,,\,\, 0 \leq i \leq k_2-1, \label{eq:def:beta}
\end{align}
and let $\Gamma= \{\gamma_i| 0 \leq i \leq |\Gamma|-1 \} := \{\alpha_i| 0\leq i \leq k_1-1\}
\cup \{\beta_i| 0 \leq i \leq k_2-1\}$ be a set with the elements indexed in
an ascending order. Here $|\Gamma | \leq k_1+k_2-1$ 
(as $\alpha_{k_1-1}=\beta_{k_2-1}=1$).
For clarity, these are illustrated in Figure~\ref{fig:rate:assign}, where
$H_1 \in \{\hao,\hbo,\hco \} $ and $H_2 \in \{\hatt,\hbt,\hct\}$.
Note that $\alpha_i,\,\, 0\leq i \leq k_1-1$ are the horizontal levels
in the plot of $\psi_1$ (see Fig.~\ref{fig:rate:assign}) which partition the 
interval $(0,1]$. 
The elements of $\Gamma$ form a partition of $(0,1]$ into $|\Gamma|$ segments. 
This is illustrated in Fig.~\ref{fig:rate:assign} for two CDFs, where 
the elements $\gamma_i$ are shown as the  levels on the $y-$axis. 
Clearly $\gamma_0=\alpha_0, \gamma_1=\beta_0,
\gamma_2=\beta_1, \gamma_3=\alpha_1, \gamma_4=1$ 
in Fig~\ref{fig:rate:assign}. 

\begin{remark} \label{rem:hor:cut}
We will often refer to the $\Gamma$ defined above as the horizontal cuts
of the CDF, in reference to Figure~\ref{fig:rate:assign}.
\end{remark}

Now for $j=1,2$, let us define the same number of `expanded' channel states of 
both the users by repeating their individual channel states appropriately
using the inverse CDF of the fading states at $\gamma_i;0 \leq i \leq |\Gamma|-1$:
\begin{align} \label{eq:disc:cdf}
h_{ji} := h_j(\gamma_i) & = \sup\{h| \psi_j(h) < \gamma_i\}.
\end{align}
For example,  in Fig.~\ref{fig:rate:assign},
the values of the `expanded' fading states $h_{1i}, 0 \leq i \leq 4$ 
of the first user are  $(\hao, \hbo, \hbo, \hbo, \hco)$ and 
the expanded states $h_{2i}, 0 \leq i \leq 4$ of the second user 
are $(\hatt, \hatt, \hbt, \hct, \hct)$.
By definition, for any $j$, the values  $h_{ji}$ are non-decreasing with
$i$ for $0 \leq i \leq 4$ .

%

We now state our main result for discrete fading states, a rate allocation 
in terms of the expanded fading states $h_{ij}$.

\begin{theorem} \label{th:discrete}
For any $\rho$ in the positive interval $\left[\frac{1}{2}\log (1+ \frac{h_{10}^2P_1}{1+h_{20}^2P_2}),
\frac{1}{2}\log (1+ h_{10}^2P_1)\right]$, the rate-strategy given by
\begin{align}
&R_1(h_{10})=\rho  \label{eq:th1a}\\ 
&R_2(h_{2i})=\frac{1}{2}\log(1+h_{1i}^2P_1+h_{2i}^2P_2)-R_1(h_{1i}) \label{eq:th1b} \\ 
&R_1(h_{1j})=\frac{1}{2}\log(1+h_{1j}^2P_1+h_{2(j-1)}^2P_2)-R_2(h_{2(j-1)}), \label{eq:th1c}
\end{align}
where $0\leq i< k$, $1\leq j< k$ and $k=k_1+k_2-1$, is outage-free and 
achieves the adaptive sum-capacity $C_{sum}(1,1,\psi_1,\psi_2)$.
\end{theorem}
In the above theorem, the rates for the users are assigned iteratively,
alternating between the users. More precisely, they are assigned to $h_{ji}$
in the lexicographic order of the pair $(i,j)$. At any stage of rate assignment,
the sum-rate is maximized with the last state (of the other user), thus
guaranteeing the maximum sum-rate in all pairs of consecutive states in this
order of the rate assignment. Though the expanded states $h_{ji}$ repeat,
it is easy to see that the mentioned rate assignment is still well defined.
That is, if $h_{ji}=h_{jk}$ for some $j,i,k$, the rate assignment 
algorithm in Theorem~\ref{th:discrete} ensures $R_j(h_{ji})=R_j(h_{jk})$.
Note that the choice of the parameter $\rho$ leaves some flexibility in the
optimal rate assignment. If either $h_{10}=0$ or $h_{20}=0$, then
$\rho$ is confined to take a single value.

The sequence of rate assignment is illustrated in 
Figure~\ref{fig:rate:assign} for two example CDFs $\psi_1(h)$ and $\psi_2(h)$.
The iterative rate-assignment is shown at the right, where
the rate-choice at the base of each arrow determines
the rate for the state at the head/front of the arrow. For example, 
the rate-choice $R_2(\hbt)$ as well as $R_2(\hct)$ are determined by 
the choice of $R_1(\hbo)$, that is, the assignment ensures that the 
rate-pairs $(\hbo, \hbt)$ and $(\hbo, \hct)$ achieve the respective 
maximum sum-rates. Similarly, $R_1(\hco)$ is determined by the choice 
made for $R_2(\hct)$.
%
%

Before proving Theorem~\ref{th:discrete}, we first provide
two alternate forms of the rate-assignment.
Th first
alternate inductive form is as follows. For any $\rho \in \left[\frac{1}{2}\log (1+ 
\frac{h_{10}^2P_1}{1+h_{20}^2P_2}), \frac{1}{2}\log (1+ h_{10}^2P_1)\right]$,
\begin{subequations}\label{eq:th1alternate}
\begin{align}
&R_1(h_{10})=\rho,\phantom{xxxx}  R_2(h_{20})=\frac{1}{2}\log (1+ h_{10}^2P_1
+h_{20}^2P_2)-\rho \\ 
&R_1(h_{1i})=R_1(h_{1(i-1)})+\frac{1}{2}\log(1+h_{1i}^2P_1+h_{2(i-1)}^2P_2)
-\frac{1}{2}\log(1+h_{1(i-1)}^2P_1+h_{2(i-1)}^2P_2), \\
&R_2(h_{2i})=R_2(h_{2(i-1)})+ \frac{1}{2}\log(1+h_{1i}^2P_1+h_{2i}^2P_2) - \frac{1}{2}\log(1+h_{1i}^2P_1+h_{2(i-1)}^2P_2)
\end{align}
\end{subequations}
for $i\geq 1$.
For a given $\rho$, this assignment can also be expressed in closed form as
\begin{subequations}\label{eq:th1closed1}
\begin{align}
&R_1(h_{10})=\rho,\phantom{xxxx}  R_2(h_{20})=\frac{1}{2}\log (1+ h_{10}^2P_1
+h_{20}^2P_2)-\rho \\ 
&R_1(h_{1i})= R_1(h_{10}) + \sum_{j=1}^i \left(\frac{1}{2}\log(1+h_{1j}^2P_1+h_{2(j-1)}^2P_2)
-\frac{1}{2}\log(1+h_{1(j-1)}^2P_1+h_{2(j-1)}^2P_2)\right), \\
&R_2(h_{2i})= 
R_2(h_{20}) +
\sum_{j=1}^i \left(\frac{1}{2}\log(1+h_{1j}^2P_1+h_{2j}^2P_2) - \frac{1}{2}\log(1+h_{1j}^2P_1+h_{2(j-1)}^2P_2)\right) 
\end{align}
\end{subequations}

\begin{remark} \label{rem:disc1}
The rate allocation in Theorem~\ref{th:discrete} is stated in a simplified, but
somewhat specialized, manner to avoid cumbersome presentation. For each
$i$, first the rate $R_1(h_{1i})$ is chosen to be the maximum possible
without violating the outage condition with the $h_{2(i-1)}$ state, and
then the maximum rate for $h_{2i}$ is chosen without violating the outage
condition with $h_{1i}$. This gives more priority to the first user.
If the priority to the second user is desired, then the order of allocation
can be the opposite without affecting the expected sum-rate. More generally,
independently for each $i$, the rates for $h_{1i}$ and $h_{2i}$ can be 
allocated in an arbitrary order. Even more generally, for each $i$,
the rates $R_1(h_{1i})$ and $R_2(h_{2i})$ can be chosen inductively
from the dominant face of a pentagon, i.e., satisfying
\begin{subequations} \label{eq:rate:rem}
\begin{align}
R_1(h_{1i}) + R_2(h_{2i}) & = \frac{1}{2}\log(1+h_{1i}^2P_1+h_{2i}^2P_2)\\
R_1(h_{1i}) & \leq \frac{1}{2}\log(1+h_{1i}^2P_1+h_{2(i-1)}^2P_2) - R_2(h_{2(i-1)})\\
R_2(h_{2i}) & \leq \frac{1}{2}\log(1+h_{1(i-1)}^2P_1+h_{2i}^2P_2) - R_2(h_{1(i-1)})
\end{align}
\end{subequations}
It is not difficult to show  that the proof of Theorem~\ref{th:discrete} given below 
will also hold true for any rate allocation satisfying the general conditions
stated in \eqref{eq:rate:rem}.
\end{remark}
We will first show that the rates given in Theorem~\ref{th:discrete} is outage-free (see
Definition~2). The next lemma will provide a building-block for the  proof. 
\begin{lemma} \label{lem:discrete}
Let $h_1$ and $h^{\prime}_1 \geq h_1$ be two channel states of user 1, and let
$h_2$ and $h_2^\prime \geq h_2$ be two channel states of user 2.
If 
\begin{align*}
 R_1(h_1) + R_2(h_2) &\leq \frac{1}{2}\log(1+h_1^2P_1+h_2^2P_2), \\
 R_1(h^{\prime}_1) + R_2(h^{\prime}_2) &\leq \frac{1}{2}\log(1+h^{\prime2}_1P_1+h^{\prime2}_2P_2),
 \\ 
\mbox{and } R_1(h_1)+R_2(h_2^\prime) &= \frac{1}{2}\log(1+h_1^2P_1+h_2^{\prime 2}P_2), 
\end{align*}
then
\begin{align*}
R_1(h^{\prime}_1)+R_2(h_2) \leq \frac{1}{2}\log(1+h^{\prime2}_1P_1+h_2^2P_2).
\end{align*}
\end{lemma}
\begin{IEEEproof} For the fading states given in the statement of the lemma, 
\begin{align}
R_1(h^{\prime}_1)+R_2(h_2) 
 &= (R_1(h_1)+R_2(h_2)) + (R_1(h^{\prime}_1)+R_2(h_2^\prime)) - (R_1(h_1)+R_2(h_2^\prime)) \nonumber \\
& \leq \frac{1}{2}\log(1+h_1^2P_1+h_2^2P_2) + \frac{1}{2}\log(1+h^{\prime2}_1P_1+h_2^{\prime 2}P_2) - \frac{1}{2}\log(1+h_1^2P_1+h_2^{\prime 2}P_2). \label{eq:lemma1}
\end{align}
Now, let us denote $u_1 = h_1^2P_1+h_2^{2}P_2$,
$u_2 = h_2^2P_1+h_2^{\prime 2}P_2$, $u_3 = h_1^2P_1+h_2^{\prime 2}P_2$,
and $u_4 = h_2^2P_1+h_2^{2}P_2$. By the hypothesis, $u_1 \leq u_i \leq u_2, i=3,4$.
Then, by Lemma~\ref{lem:log:concave}, we have
\begin{align}
\frac{1}{2}\log(1+ u_1) + \frac{1}{2}\log(1+u_2) \leq \frac{1}{2}\log(1+u_3) 
	+ \frac{1}{2}\log(1+ u_4). \label{eq:lemm:5:pf} 
\end{align}
The lemma is proved by applying \eqref{eq:lemm:5:pf} to (\ref{eq:lemma1}).
\end{IEEEproof}

{\it Proof of Theorem~\ref{th:discrete}:}
In order to check that a given  rate-strategy is outage-free, we need
to verify three constraints of the pentagon for each pair of states.
Let us first check the sum-rate constraint, followed by the individual
rate constraints.

 Let $h$ and $\h2$ be arbitrary states of user 1 and user 2 respectively.
We will show that the chosen rate-pair is inside the corresponding 
MAC pentagon, and thus outage-free.
By the definition in \eqref{eq:disc:cdf}, for some $i$ and $j$, $h = h_{1i}$ and $\h2 = h_{2j}$. 
To check the sum-rate constraint of \eqref{eq:mac:pent},
let us assume w.l.o.g that $i\leq j$. The proof will be done by induction 
on $|j-i|$. If $i=j$, then by (\ref{eq:th1b}), $R_1(h) + R_2(\h2)
= \frac{1}{2} \log (1+ h^2P_1 + {\h2}^{ 2}P_2)$. For $j=i+1$,
the maximum sum-rate is achieved for state-pairs $(h_{1i},h_{2i}),
(h_{1j},h_{2i})$, and $(h_{1j},h_{2j})$ by \eqref{eq:th1b}, \eqref{eq:th1c} and \eqref{eq:th1b}
respectively. So, Lemma~\ref{lem:discrete} gives $R_1(h_{1i}) + R_2(h_{2j})
\leq \frac{1}{2} \log (1+ h_{1i}^2P_1 + {h_{2j}}^{ 2}P_2)$.
Now suppose for some $t\geq 2$, and all $i,j$ with $|j- i|<t$, it holds that
$R_{2}(h_{2j}) + R_{1}(h_{1i}) \leq \frac{1}{2}\log(1+ h_{1i}^2P_1+h_{2j}^2P_2)$. 
Then for $j=i+t$, we have
\begin{subequations}
\begin{align}
R_{2}(h_{2(j-1)}) + R_{1}(h_{1i}) & \leq \frac{1}{2}\log(1+ h_{1i}^2P_1+h_{2(j-1)}^2P_2) \label{eq:ind1}\\
R_{2}(h_{2(j-1)}) + R_{1}(h_{1(j-1)}) & = \frac{1}{2}\log(1+ h_{1(j-1)}^2P_1+h_{2(j-1)}^2P_2) \label{eq:ind2}\\
R_{2}(h_{2j)}) + R_{1}(h_{1(j-1)}) & \leq \frac{1}{2}\log(1+ h_{1(j-1)}^2P_1+h_{2j}^2P_2). \label{eq:ind3}
\end{align}
\end{subequations}
where \eqref{eq:ind2} follows from \eqref{eq:th1b}, and \eqref{eq:ind1} and 
\eqref{eq:ind3} follow from the induction hypothesis.
Using this in 
Lemma~\ref{lem:discrete}, it follows that $R_{2}(h_{2j}) + R_{1}(h_{1i}) 
\leq \frac{1}{2}\log(1+ h_{1i}^2P_1+h_{2j}^2P_2)$. 
This completes the proof by induction. 


Having verified the sum-rate constraint, let us also prove $R_j(h_{ji}) \leq \frac{1}{2} 
\log(1+ h_{ji}^2 P_j)$, for $j=1,2$. We do this by induction on $i$. 
The base case of $i=0$ follows from \eqref{eq:th1a} and \eqref{eq:th1b}.
Now let us consider $i>0$. We give the proof for $j=1$, and the proof
for $j=2$ follows similarly.
By (\ref{eq:th1b}) and (\ref{eq:th1c}),
\begin{align}
R_1(h_{1i})  
& = \frac{1}{2}\log(1+h_{1i}^2P_1+h_{2(i-1)}^2P_2)  
 - \frac{1}{2}\log(1+h_{1(i-1)}^2P_1+h_{2(i-1)}^2P_2)+R_1(h_{1(i-1)})  \nonumber \\
& \leq \frac{1}{2}\log(1+h_{1i}^2P_1+h_{2(i-1)}^2P_2)
	 + \frac{1}{2}\log(1+h_{1(i-1)}^2P_1) 
 - \frac{1}{2}\log(1+h_{1(i-1)}^2P_1+h_{2(i-1)}^2P_2) \label{eq:thpf1} \\
& = \frac{1}{2}\log(1+h_{1i}^2P_1+h_{2(i-1)}^2P_2)  
 - \frac{1}{2}\log\left(1+\frac{h_{2(i-1)}^2P_2}{1+h_{1(i-1)}^2P_1}\right) \nonumber \\
& \leq \frac{1}{2}\log(1+h_{1i}^2P_1+h_{2(i-1)}^2P_2)  
 - \frac{1}{2}\log\left(1+\frac{h_{2(i-1)}^2P_2}{1+h_{1i}^2P_1}\right) \label{eq:thpf2} \\
& = \frac{1}{2}\log(1+h_{1i}^2P_1). \nonumber
\end{align}
Inequality (\ref{eq:thpf1}) follows from the induction hypothesis, whereas
(\ref{eq:thpf2}) uses the fact that $h_{1i} \geq h_{1(i-1)}$.

Let us now prove that our rate-strategy maximizes the expected sum-rate.
The key is to notice that, using the inverse CDF definitions of \eqref{eq:cdf:inv}, 
our rate-allocation
ensures that for any $x\in [0,1)$, $R_1(h_1(x)) + R_2(h_2(x))
= \frac{1}{2}\log (1+h_1^2(x)P_1 + h_2^2(x)P_2)$. But any outage-free
rate-allocation $(R_1(\cdot), R_2(\cdot))$ satisfies
\begin{align} 
\eE(R_1(H_1)) + \eE(R_2(H_2)) 
& = \int_0^1 (R_1(h_1(x)) + R_2(h_2(x))) dx \nonumber \\
& \leq \frac{1}{2} \int_0^1 \log (1+h_1^2(x)P_1 + h_2^2(x)P_2) dx. \label{eq:thpf3} 
\end{align}
The equality in the first line is by \eqref{eq:rate:cdf}, and the 
inequality above follows from \eqref{eq:mac:pent}.
Clearly, the proposed scheme achieves this upper bound and
this completes the proof of the theorem.\hfill{\rule{2mm}{2mm}}

\subsection{DISCRETE FADING STATES: MULTIPLE USERS}
\label{subsec:dis:mult}
The results from the previous sections can be extended to multiple users.
We first  discuss the rate-allocation achieving the adaptive sum-capacity  
for arbitrary discrete states for each user. 
Let user $i, ~1 \leq i \leq N$ have $k_i$ channel states
with probabilities $p_{ij}; 0\leq j \leq k_i-1$. Let us denote
the CDF values of the channels as
\begin{align*}
& \alpha_{il} = \sum_{j=0}^l p_{ij}\,,\,\, 0 \leq l \leq  k_i-1, \\ 
\end{align*}
and let $\{\gamma_l| 0\leq l \leq |\Gamma|-1\} = \cup_{i} \{\alpha_{il}| 0\leq l < k_i\}$ 
be these values indexed in the ascending order, where  $|\Gamma| \leq \sum_{i=1}^N k_i  - l + 1$.
These definitions are analogous to the
two user ones in \eqref{eq:def:alpha}.
The values 
$\alpha_{il},\,\, 0\leq l \leq k_i-1$ are all the
horizontal levels in the interval $(0,1]$ in the discrete CDF $\psi_i$.
 $\gamma_l,\,\,0\leq l < |\Gamma|$ denote the union of these horizontal levels. 

Now for $1 \leq i \leq N$, let us define the same number of `expanded' channel states of 
the users by repeating their individual channel states appropriately
using the inverse CDF of the fading states at $\gamma_l;0\leq l\leq |\Gamma|-1$:
\begin{align} \label{eq:disc:cdfmult}
h_{il} := h_i(\gamma_l) & = \sup\{h| \psi_i(h) < \gamma_l\}.
\end{align}
In this notation, $h_{i0}$ denotes the fading state of lowest magnitude for user $i$.

Now, we state the result for discrete fading states. The empty sum
is defined to be zero as usual.
\begin{theorem} \label{th:discretemult}
Let $R_i(h_{i0}), 1 \leq i \leq N$ be such that for any $S \subset \{ 1,2,\cdots, N\}$
\begin{align}
 \sum_{i \in S}R_i(h_{i0}) &\leq \frac12 \log(1+\sum_{i \in S}h_{i0}^2P_i) \mbox{ and } \label{eq:th:disc:mult1} \\
 \sum_{i=1}^N R_i(h_{i0}) &= \frac12 \log(1+\sum_{i=1}^N h_{i0}^2P_i) \label{eq:rate:mult:init}.
\end{align}
Then, the inductive rate allocation given by
\begin{align} \label{eq:dis:mult:rate}
&R_i(h_{il})=\frac{1}{2}\log \left(1+\sum_{k=1}^i h_{kl}^2P_k+\sum_{j=i+1}^Nh_{j(l-1)}^2P_j\right)-\left( \sum_{k=1}^{i-1} R_k(h_{kl}) \right)- \left( \sum_{j=i+1}^N R_j(h_{j(l-1)})\right), ~ 1 \leq i \leq N, 
\end{align}
where $1\leq l< |\Gamma|$, is an
outage-free strategy achieving the adaptive sum-capacity.
\end{theorem}

\begin{remark} \label{rem:mult1}
The above rate allocation can also be expressed in an alternate
form similar to \eqref{eq:th1alternate}, and also in closed form similar to
\eqref{eq:th1closed1}. However, as discussed in Remark~\ref{rem:disc1}, there is
a lot more flexibility in the sum-rate optimal rate allocation than what is
reflected in Theorem~\ref{th:discretemult}. For each $l$, the rate
allocation for the states $h_{1l}, h_{2l}, \cdots, h_{Nl}$ can be
done in any order while ensuring the outage-free condition with the
already rate-assigned states. Even more generally, inductively for each $l$, any
rates can be chosen for the states $h_{1l}, h_{2l}, \cdots, h_{Nl}$ as long as they satisfy
\begin{align}
\sum_{i=1}^N R_i(h_{il}) & = \frac{1}{2}\log(1+\sum_{k=1}^N h_{kl}^2P_k) \label{eqmultoutg1}\\
\sum_{i\in S} R_i(h_{il}) & \leq \frac{1}{2}\log(1+\sum_{k\not\in S} h_{k(l-1)}^2P_k +\sum_{k\in S} h_{kl}^2P_k) - \sum_{k\not\in S} R_k(h_{k(l-1)}) \phantom{xxx} \forall S\subset \{1,2,\cdots,N\}, \label{eqmultoutg}
\end{align}
where $R_k(h_{kl}):= 0$ and $h_{kl}:=0$ for $l<0$.
\end{remark}

\begin{IEEEproof}
 Let $S \subseteq \{1,2,3, \cdots, N\}$ be a set of users. Without loss of generality, assume that the indexes in $S$
are in the ascending order. Let $h_{il_i}$ be some channel states of
these users, where $1\leq l_i\leq |\Gamma|-1$ for all $i$. To ensure the outage-free conditions, we will show that 
\begin{align} \label{eq:dis:mult:ind1}
 \sum_{i \in S}R_i(h_{il_i})\leq \frac12 \log\left(1+\sum_{i \in S}h_{il_i}^2P_i\right).
\end{align}
Recall that the fading state $h_{il_i}$ for user~$i$ is defined 
by~\eqref{eq:disc:cdfmult}. 
Notice that in the special case where $S=\{1,2,\cdots, N\}$, and $l_j=l, 
\forall j\in S$, the rate-allocation in \eqref{eq:dis:mult:rate}  guarantees 
(by taking $i=N$) a rate-tuple on the dominant face of the capacity-region.
Let
us consider the ordered-pair $(l_i,i)$. Let $(l_k,k)$ be the highest pair in the 
lexicographical ordering of the states over all the users in $S$. That is, 
if $l= \max \{l_i:i\in S\}$, then $k=\max\{i\in S: l_i=l\}$.


%
%
%
Let us define, for any $t$, $S_1[t] := \{1,\cdots, t\} \bigcap S$ and $S_2[t] := \{t, \cdots, N\}\bigcap S$. 
Using \eqref{eq:dis:mult:rate} 
\begin{align}
\sum_{i \in S} R_i(h_{il_i}) &=  \ssuma{1}{k-1} R_j(h_{jl_j}) + R_k(h_{kl_k}) + \ssumb{k+1}{N} R_j(h_{jl_j}) \\
	&= \ssuma{1}{k-1} R_j(h_{jl_j})  +
	\frac{1}{2}\log \left(1+\sum_{j=1}^k h_{jl_k}^2P_j+\sum_{j=k+1}^N h_{j(l_k-1)}^2P_j\right)  \notag \\
	&\phantom{wwwwwwwwww}- \sum_{j=1}^{k-1} R_j(h_{j{l_k}}) 
		- \sum_{j=k+1}^N R_j(h_{j(l_{k}-1)})
		+ \ssumb{k+1}{N} R_j(h_{jl_j}) \notag\\
	&= \ssuma{1}{k-1} R_j(h_{jl_j}) +  
		\frac{1}{2}\log \left(1+\sum_{j=1}^k h_{jl_k}^2P_j+\sum_{j=k+1}^N h_{j(l_k-1)}^2P_j\right)\notag  \\
	&\phantom{ww} - \sum_{j=1}^{k-1} R_j(h_{j{l_k}}) - \sum_{j=k}^N R_j(h_{j(l_{k}-1)})
		+ R_k(h_{k(l_k-1)}) + \ssumb{k+1}{N} R_j(h_{jl_j}) \notag  \\
	&= \ssuma{1}{k-1} R_j(h_{jl_j}) + R_k(h_{k(l_k-1)}) + \ssumb{k+1}{N} R_j(h_{jl_j})
		 - \sum_{j=1}^{k-1} R_j(h_{j{l_k}}) - \sum_{j=k}^N R_j(h_{j(l_{k}-1)}) \notag \\
		&\phantom{wwww}+ \frac{1}{2}\log \left(1+\sum_{j=1}^k h_{jl_k}^2P_j+\sum_{j=k+1}^N h_{j(l_k-1)}^2P_j\right).
	\label{eq:si:proof:3}
\end{align}

We now provide an inductive argument to show \eqref{eq:dis:mult:ind1}. 
The base case for any $S$ and $l_i=0,\,\forall i\in S$ holds by \eqref{eq:th:disc:mult1}.
We now assume that \eqref{eq:dis:mult:ind1} is true for all 
$(S, (l_i)_{i\in S})$ with strictly lower $(l_k, k)$. In particular,
we assume that
\begin{align} \label{eq:sec3b:lem:1}
\ssuma{1}{k-1} R_j(h_{jl_j}) + R_k (h_{k({l_k}-1)}) + \ssumb{k+1}{N} R_j(h_{jl_j})\leq
                \frac 12 \log \left(1 + \ssuma{1}{k-1} h_{jl_j}^2 P_j +  h_{k(l_k-1)}^2 P_k
                        +  \ssumb{k+1}{N} h_{jl_j}^2 P_j \right).
\end{align}
By \eqref{eq:dis:mult:rate}, we have
\begin{align} \label{eq:sec3b:ass:1}
\sum_{j=1}^{k-1} R_j(h_{j{l_k}}) + \sum_{j=k}^N R_j(h_{j(l_{k}-1)})
        = \frac{1}{2}\log \left(1+\sum_{j=1}^{k-1} h_{jl_k}^2P_j+\sum_{j=k}^N h_{j(l_k-1)}^2P_j\right).
\end{align}
Using Lemma~\ref{lem:log:concave}, since the arguments of the logarithm 
sum to the same  on both sides, we also have
\begin{multline}
\frac 12 \log \left(1 + \ssuma{1}{k-1} h_{jl_j}^2 P_j +  h_{k(l_k-1)}^2 P_k
                        +  \ssumb{k+1}{N} h_{jl_j}^2 P_j \right)
        +\frac{1}{2}\log \left(1+\sum_{j=1}^k h_{jl_k}^2P_j+\sum_{j=k+1}^N h_{j(l_k-1)}^2P_j\right) \\
        \leq \frac 12 \log \left(1 + \ssuma{1}{N} h_{jl_j}^2 P_j \right)
        +\frac{1}{2}\log \left(1+\sum_{j=1}^{k-1} h_{jl_k}^2P_j+\sum_{j=k}^N h_{j(l_k-1)}^2P_j\right).
\label{eq:sec3b:complicated}
\end{multline}
Now, using \eqref{eq:sec3b:lem:1}, \eqref{eq:sec3b:ass:1}, and \eqref{eq:sec3b:complicated} in \eqref{eq:si:proof:3}, we have the result,
that is, \eqref{eq:dis:mult:ind1}.

To complete the proof, we need to check that the rate-allocation is optimal. This follows as in \eqref{eq:thpf3}, since
we have ensured equality to the maximal sum-rate for every horizontal cut (see Remark~\ref{rem:hor:cut})
of the CDFs. 

\end{IEEEproof}

\subsection{CONTINUOUS VALUED FADING STATES}
\label{subsec:cont}

When the fading coefficients take continuous values, the rate-allocation
algorithm developed in the last section cannot be applied directly.
However, one can discretize the channel states with as small a step size
as desired and then use the rate-allocation algorithm. This is expected
to give a near-optimal rate-allocation. 
In the limit where the discrete step-size approaches zero, the algorithm
provides a closed form
elegant solution (Theorem~\ref{thm:cont:capa} below) to the optimal rate-allocation. 
Apart from its technical merit,
the explicit rate allocation is widely useful, since continuous-valued
distributions like Rayleigh are commonly used to model wireless links. 
Here we will directly 
provide the rate-allocation formula and prove that it is outage-free
and sum-rate optimal.
We delegate the details of how the closed form expression
was obtained from the algorithm in Theorem~\ref{th:discrete} to 
Appendix~\ref{disc2cont}.  Our results are true for a wide class of 
distributions including
combinations of continuous valued and discrete states.

Consider two continuous valued fading distributions $\psi_1(h)$ and 
$\psi_2(h)$. 
Recall that $h_j(x)=\psi_j^{-1}(x)$
is the inverse CDF of user~$j$, as defined in \eqref{eq:cdf:inv}.

%
\begin{theorem}\label{thm:cont:capa}
For a two user Gaussian MAC with  fading distributions $\psi_1(\cdot)$ and $\psi_2(\cdot)$, 
and with respective 
transmit powers $P_1$ and $P_2$, the adaptive sum-capacity $C_{sum}(1,1,\psi_1,\psi_2)$
with individual CSI is achieved by the rate-allocation
\begin{align} \label{eq:rate:cont}
R_i(h) = R_i(\mi)+\int_{\mia}^h \dfrac{yP_i}{1+y^2P_i + \sum_{j\neq i}(\psi_j^{-1}(\psi_i(y)))^2P_{j}}\,\mathrm{d}y ,~h\geq h_i(0), ~i\in \{1,2\},
\end{align}
for any $R_1(\mione), R_2(\mitwo)$ satisfying 
\begin{align} 
R_i(\mi) &\leq \frac 12 \log(1+h_i^2(0)P_i) ,~i\in \{1,2\} \notag  \\
\sum_{i=1}^2R_i(\mi) &= \frac 12 \log(1+h_1^2(0)P_1+h_2^2(0)P_2). \notag
\end{align}
\end{theorem}

\begin{IEEEproof}
Let us first find an upper bound for the expected sum-rate of any achievable scheme.
\begin{align}
\sum_{i=1}^2 E[R_i(H_i)] = 
	\int_0^\infty R_1(h) d\psi_1(h)\, +\int_0^\infty R_2(h) d\psi_2(h). \notag
\end{align}
By the same steps as the discrete-state derivation in \eqref{eq:thpf3},
\begin{align} \label{eq:ub4}
\sum_{i=1}^2 E[R_i(H_i)] 
	\leq \int\limits_0^1 \frac 12 \log(1+h_1^2(x)P_1+h_2^2(x)P_2)\,dx 
\end{align}

To complete the proof, we will  show 
that the rate allocation in 
\eqref{eq:rate:cont} is outage free and it achieves the upper bound in 
\eqref{eq:ub4}.
\begin{claim} \label{claim:one}
The rate allocation given in \eqref{eq:rate:cont} is outage-free.
\end{claim}
\begin{IEEEproof}
For the rate functions in \eqref{eq:rate:cont}, we will show that 
$\forall (h_1, h_2)$ such that $h_i \geq \mia$, $i=1,2$,
\begin{align*}
 R_1(h_1) + R_2(h_2) &\leq \frac 12 \log(1+h_1^2 P_1 + h_2^2 P_2 ),\\
\mbox{and }R_i(h_i)&\leq \frac 12 \log(1+h_i^2P_i),~ i\in{1,2}. 
\end{align*}
Showing this require a bit of calculus, and is relegated to
Appendix~\ref{app:sec3c}.
\end{IEEEproof}
Let us now show the optimalty of the allocation in \eqref{eq:rate:cont}.
\begin{lemma} \label{lem:two}
For $x\in[0,1]$ and the rate allocation in \eqref{eq:rate:cont}, 
$$
R_1(h_1(x)) + R_2(h_2(x)) = \frac 12 \log(1+ h_1^2(x)P_1 + h_2^2(x)P_2).
$$
\end{lemma}
\begin{IEEEproof}
From the rate allocation in \eqref{eq:rate:cont}, it follows that
$$R_1(h_1(0)) + R_2(h_2(0)) = \frac 12 \log(1+ h_1^2(0)P_1 + h_2^2(0)P_2).$$
Also, for $x >0$, 
\begin{align}
\sum_{i=1}^2R_i(h_i(x)) =
	R_1(h_1(0)) + R_2(h_2(0))+\int\limits_{h_1(0)}^{h_1(x)} 
	\dfrac{yP_1}{1+y^2P_1+(\psi_2\textsuperscript{-1}(\psi_1(y)))^2P_2}\,dy
	+\int\limits_{h_2(0)}^{h_2(x)} 
	\dfrac{yP_2}{1+y^2P_2+(\psi_1\textsuperscript{-1}(\psi_2(y)))^2P_1}\,dy 
\end{align}
Substituting $\psi_1\textsuperscript{-1}(\psi_2(y))=z$ in the second integral,
we get
\begin{align*}
\sum_{i=1}^2R_i(h_i(x)) &=
	R_1(h_1(0)) + R_2(h_2(0))+\int\limits_{h_1(0)}^{h_1(x)} 
	\dfrac{yP_1}{1+y^2P_1+(\psi_2\textsuperscript{-1}(\psi_1(y)))^2P_2}\,dy  +\int\limits_{h_1(0)}^{h_1(x)} \dfrac{P_2\psi_2\textsuperscript{-1}(\psi_1(z))(\psi_2\textsuperscript{-1}(\psi_1(z))' }{1+z^2P_1+(\psi_2\textsuperscript{-1}(\psi_1(z)))^2P_2}\,dz\\
	&=R_1(h_1(0)) + R_2(h_2(0))+\int\limits_{h_1(0)}^{h_1(x)} \dfrac{P_2\psi_2\textsuperscript{-1}(\psi_1(z))(\psi_2\textsuperscript{-1}(\psi_1(z))'+zP_1 }{1+z^2P_1+(\psi_2\textsuperscript{-1}(\psi_1(z)))^2P_2}\, dz \\
	&=R_1(h_1(0)) + R_2(h_2(0))+\int\limits_{1+h_1^2(0)P_1 + h_2^2(0)P_2}^{1+h_1^2(x)P_1+h_2^2(x)P_2} \dfrac{1}{2p}\,dp \phantom{xxxx} (\text{by substituting } 1+z^2P_1+(\psi_2\textsuperscript{-1}(\psi_1(z)))^2P_2 = p )\\
	&=\frac 12 \log(1+ h_1^2(0)P_1 + h_2^2(0)P_2)+\dfrac{1}{2}\log(1+h_1^2(x)P_1+h_2^2(x)P_2)-\frac 12 \log(1+ h_1^2(0)P_1 + h_2^2(0)P_2) \\
	&=\dfrac{1}{2}\log(1+h_1^2(x)P_1+h_2^2(x)P_2).
\end{align*}
This proves the lemma.
\end{IEEEproof}
We have thus shown that the rate allocation in \eqref{eq:rate:cont} is optimal for 
achieving the adaptive sum-capacity. This completes the proof of  Theorem~\ref{thm:cont:capa}.
\end{IEEEproof}

The rate-allocation in Theorem~\ref{thm:cont:capa} reduces to the
optimal rate-allocation formula~\eqref{eq:th1closed1} for the discrete fading 
states as a special case (with $\rho = R_1(h_1(0))$). The formula
also extends to more users than two, presented in the next subsection.

\subsection{CONTINUOUS CASE: MULTIPLE USERS}

\label{subsec:cont:mult}

For $N$ users with continuous valued fading states, the rate allocation in \eqref{eq:rate:cont} is generalized in the following
theorem.
\begin{theorem}\label{thm:cont:mult}
The rate allocation given by:
\begin{align} \label{eq:adptcpmult}
R_i(h) = R_i(\mi)+\int_{\mia}^h \dfrac{yP_i}{1+y^2P_i + \sum_{j\neq i}(\psi_{j}^{-1}(\psi_i(y)))^2P_{j}}\,\mathrm{d}y ,~h\geq \mia, ~1\leq
i \leq N,
\end{align}
for any $R_i(\mi)$ satisfying
\begin{align} 
\sum_{i\in S} R_i(\mi) \leq \frac 12 \log(1+\sum_{i\in S}h_i^2(0)P_i), ~S \subset \{1,2,\cdots,N\}, \notag
\end{align}
%
achieves the adaptive sum-capacity $C_{sum}(\mathbf 1,  \bpsi)$ of an $N-$user individual CSI MAC,
where $\mathbf 1$ is a vector of all ones.
\end{theorem}
The proof of this theorem is similar to the two-users case, and
is relegated to Appendix~\ref{sec:multderive}


\subsection{SIMULATION STUDY}
\begin{figure}[htbp]
\begin{center}
\includegraphics[scale=0.6]{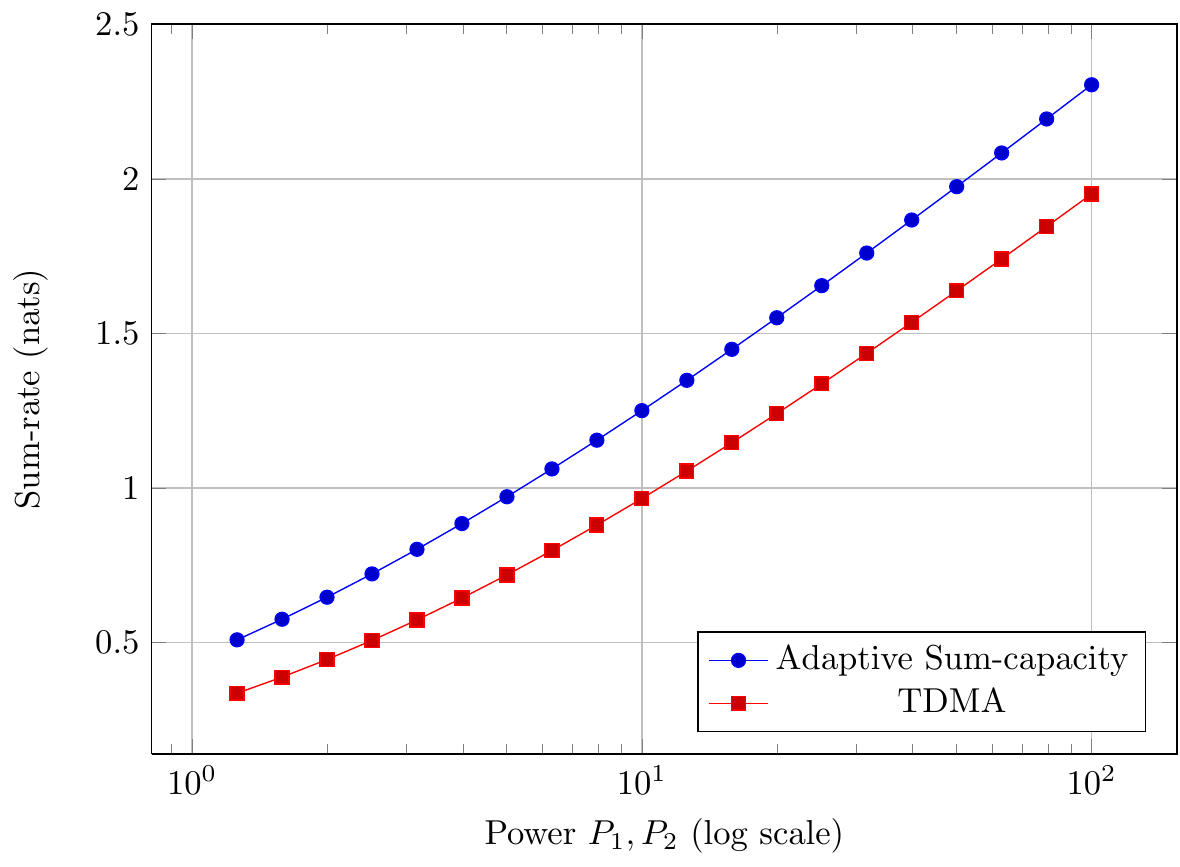}
\caption{Two User Adaptive Sum-capacity with $P_1=P_2$ for independent Rayleigh and uniform fading states\label{fig:plot:asum}}
\end{center}
\end{figure}

  We demonstrate the advantage of our solution by an example.
Let $\psi_1(h)$ be the normalized Rayleigh CDF, and $\psi_2(h)$ be
uniformly distribution in $[0,\sqrt{3}]$.  Thus $\eE|H_1|^2 = \eE|H_2|^2 = 1$. 
Figure~\ref{fig:plot:asum} shows the adaptive sum-capacity when the transmit power 
is varied while maintaining $P_1=P_2$. For comparison, we also
show the sum-rate achieved by the conventional
strategy of time division multiplexing (TDMA), where the time is divided into
equal-sized slots. The same cap on transmit power is imposed in both
cases. Clearly, the proposed solution outperforms the conventional
strategy.

An astute reader will quickly point out that a generalized TDM scheme
can also employ more power while transmitting, and still maintain the
short-term average power constraint. We can  go even one step further and employ
the best power-adaptation scheme for TDMA. Nevertheless, even in this case, we will show
in Section~\ref{sec:PCReg} that our rate-adaptation strategy will perform
significantly better. In fact, an improved performance can be achieved by
employing the TDMA power-adaptation itself, which is in general suboptimal
for achieving the power-controlled adaptive sum-capacity, this
is demonstrated in Figure~\ref{fig:tdm:pow:ctrl}.

\section{Adaptive capacity region}
\label{sec:adap:capa:reg}
Recall that the adaptive capacity region is the collection of all rate-tuples of the form, 
$(\mathbb{E}_{H_1}[R_1(H_1)],$ $\cdots,\mathbb{E}_{H_L}[R_L(H_L)])$, 
where the rate-allocation strategies  do not lead to outage in any block.  
The adaptive capacity region in the presence of individual
CSI can be  characterized by maximizing the weighted sum-rate 
$\sum_{i=1}^L w_i \mathbb{E}_{H_i}[R_i(H_i)]$ for all non-negative vectors $\bow$. 
For the economy of space, we present the adaptive capacity region for the case of $L=2$, extending 
to more users is reasonably straightforward. 
We also assume in this section that the transmitter $i$ uses a fixed transmit power $P_i$ 
for all fading states, i.e $P_i(h)=P_i,~ \forall h,~i=1,2$. 
The general case where power control is allowed  will be addressed in Section \ref{sec:PCReg}.

\begin{figure}[htbp]
\begin{center}
\begin{tikzpicture}[line width=1.5pt, scale=1.5]

\draw[->] (0,0) --++(4,0) node[right]{$h$};
\draw[->] (0,0) --++(0,3.5) node[above]{$x$};
\draw plot[smooth] coordinates {(0,0) (0.5,0.25) (0.75,0.75) (1,1.75) 
	(1.175,2.25) (1.3,2.5) (1.5,2.75) (1.75,2.95)  (2,3) }node [left,above,]{$\psi_2$};
\draw[red] (0,0) -- (2.5,3) node[above]{$\psi_1$};
\draw[thin] (0,3) --++(4,0);
\foreach \x in {0.25,0.5,...,3} {\draw[dashed,thin] (0,\x) --++(3.5, 0);}
\foreach \x in {0.25,0.5,0.75,1} {\node[left]  at (0,{3*\x}) {$\x$} ;}
\draw[gray, line width=2pt] (0,2.25) --++(3.5,0);

\draw[thin] (1.175,2.25) circle (0.05cm) --++(0,-2.25) node[below]{$h_2$};
\draw[thin] ({2.25*2.5/3},2.25) circle (0.05cm)--++(0,-2.25) node[below]{$h_1$};

\pgftransformxshift{5cm}
\begin{scope}[scale=1.25]
\draw[->,thin] (0.25,0.25) --++(3,0) node[right]{$R_1$};
\draw[->,thin] (0.25,0.25) --++(0,2.75) node[above]{$R_2$};

\draw[line width=1pt] (2.5,0.25) --(2.5,1) --(1.5,2)  --(0.25,2);

\draw[thin, dotted] (2.15,1.35) --++(-1.9,0) node[left,scale=0.85]{$R_1(h_1)$};
\draw[thin, dotted] (2.15,1.35) --++(0,-1.1) node[below,scale=0.85]{$R_2(h_2)$};

\draw[dashed,thin] (2.5,0.25) --node [scale=0.75, below, pos=0.35,sloped]
{$\frac 12 \log (1 + h_2^2P_2)$} ++(0,2.5); 
\draw[dashed,thin] (0.5,2) -- node [scale=0.75,pos=0.3,yshift=0.25cm]
{$\frac 12 \log (1 + h_1^2P_1)$} ++(2.5,0); 
\end{scope}
\end{tikzpicture}
\end{center}
\caption{Rate-allocation for Sum-capacity \label{fig:illus:1}}
\end{figure}
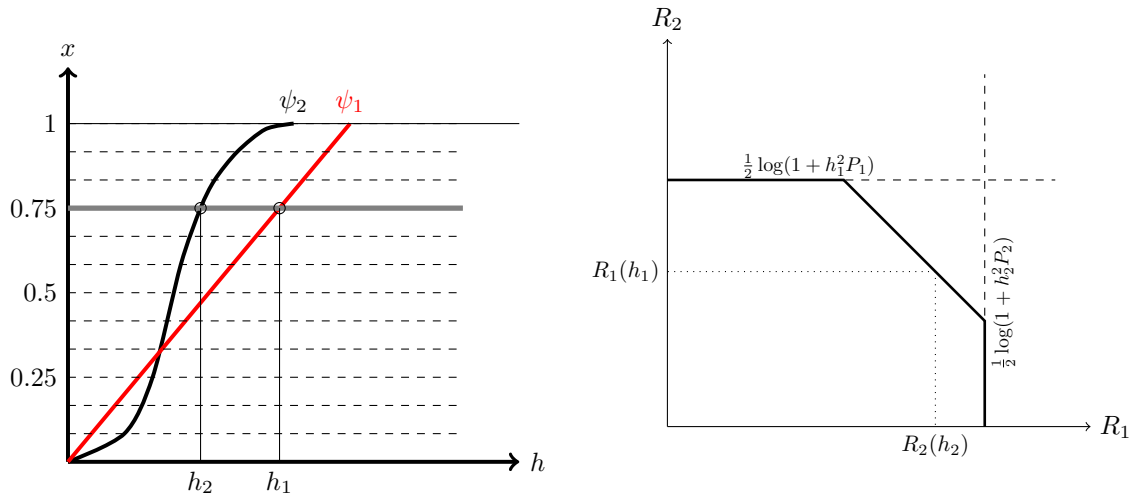

Without loss of generality, let us describe the solution for $w_1=1$
and $w_2=\alpha \leq 1$, 
the opposite case will follow by a simple renaming of the variables.
In terms of the notation in Section~\ref{sec:model} (see Definition~\ref{def:w:sum}), we have to evaluate 
$C_{sum}(1,\alpha, \psi_1, \psi_2)$, where $\psi_i, i=1,2$ are the respective CDFs of
the two links.
Using the definition of inverse in~\eqref{eq:cdf:inv}, we can write
\begin{align}
 \mathbb{E} R_1(H_1)+\alpha \mathbb{E} R_2(H_2) &= \int_0^\infty 
	\!\! R_1(h_1) d\psi_1(h_1) +\alpha\int_0^\infty \!\! R_2(h_2) d\psi_2(h_2)  \notag \\
	&= \int_0^1 \left( R_1(h_1(x)) + \alpha R_2(h_2(x))  \right)dx. \label{eq:wsum:1}
\end{align}
When $\alpha=1$, the 
sum of terms  inside the integral of \eqref{eq:wsum:1} is maximized by the corresponding sum-rate.
 This will suggest choosing a suitable operating point for every pair $(h_1(x),h_2(x))$ 
on the dominant face of the corresponding capacity pentagon.  
This concept was already explained in the example shown in Figure~\ref{fig:rate:assign} for 
discrete fading states. An analogous picture for the continuous case is shown in 
Figure~\ref{fig:illus:1} with the respective CDFs $\psi_1$ and $\psi_2$. 
For every horizontal cut there, the proposed rate-allocation chooses a point on the
dominant face of the corresponding pentagon. Figure~\ref{fig:illus:1} shows the
rate-allocation $(R_1(h_1),R_2(h_2))$ for a particular cut which corresponds to a CDF value of $0.75$.

%

\noindent
For $\alpha < 1$, a similar point-wise maximization of the weighted sum-rate
at all horizontal levels will end up choosing the right corner-point at
such state-pairs. This does not ensure outage-free operation for
state pairs $(h_1,h_2)$ for which $\psi_1(h_1) > \psi_2(h_2)$.
This is because, if $h_1^\prime$ is such that
$\psi_1(h_1^\prime) = \psi_2(h_2)$ (this means that $h_1^\prime < h_1$),
then the sum-rate at $(h_1,h_2)$ is
\begin{align*}
\frac{1}{2}\log (1+h_1^2P_1) + \frac{1}{2}\log 
\left(1+\frac{h_2^2P_2}{1+h_1^{\prime 2}P_1}\right) & >
\frac{1}{2}\log (1+h_1^2P_1) + \frac{1}{2}\log 
\left(1+\frac{h_2^2P_2}{1+h_1^{2}P_1}\right)\\
& = \frac{1}{2}\log (1+h_1^2P_1 +h_2^2P_2).
\end{align*}

\begin{figure}[htbp]
\begin{center}
\begin{tikzpicture}[scale=1.5]

\draw[->] (0,0) --++(3,0) node[right]{$h$};
\draw[->] (0,0) --++(0,3.5) node[above]{$x$};
\draw[line width=1.5pt] plot[smooth] coordinates {(0,0) (0.5,0.25) (0.75,0.75) (1,1.75) 
	(1.175,2.25) (1.3,2.5) (1.5,2.75) (1.75,2.95)  (2,3) }node [left,above,]{$\psi_2$};
\draw[red,line width=1.5pt] (0,0) -- (2.5,3) node[above]{$\psi_1$};
\draw[thin] (0,3) --++(2.5,0);
\foreach \x in {0.25,0.5,...,3} {\draw[dashed,thin,gray] (0,\x) --++(2.5, 0);}
\foreach \x in {0.25,0.5,0.75,1} {\node[left]  at (0,{3*\x}) {$\x$} ;}


\pgftransformxshift{4.5cm}
\draw[->] (0,0) --++(3,0) node[right]{$h$};
\draw[->] (0,0) --++(0,3.5) node[above]{$x$};
\draw[red, line width=1.5pt] (0,0) -- (2.5,3) node[above,xshift=0.15cm]{$\phi_1$};
\draw[thin] (0,3) --++(2.5,0);
\foreach \x in {0.25,0.5,...,3} {\draw[dashed,thin,gray] (0,\x) --++(2.5, 0);}
\foreach \x in {0.25,0.5,0.75,1} {\node[left]  at (0,{3*\x}) {$\x$} ;}
\pgftransformyshift{0.75cm}
\draw[line width=2pt] (0,-0.75) --++(0,0.75) plot[smooth,yscale=0.75] coordinates {(0,0) (0.5,0.25) (0.75,0.75) (1,1.75) 
	(1.175,2.25) (1.3,2.5) (1.5,2.75) (1.75,2.95)  (2,3) }node [left,below, yshift=-0.15cm]{$\phi_2$};
\draw[<->] (1.5,-0.75) -- node[fill=white,right]{$1-\alpha$} ++(0,0.75);
\draw[gray] (0,0) --++(1.50,0);
\end{tikzpicture}
\caption{Obtaining the Modified CDFs \label{fig:illus:2}}
\end{center}
\end{figure}
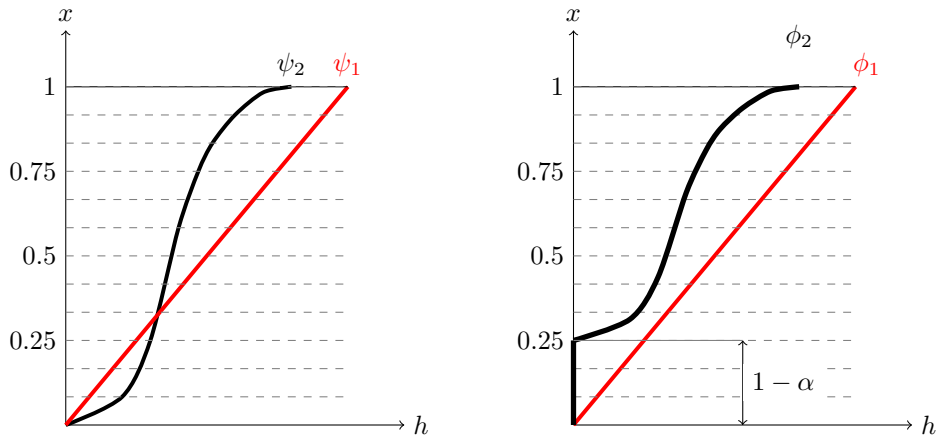

However, we will show now that the weighted sum-rate maximization problem
can be written as an equivalent sum-rate maximization problem over
a new channel state-distribution for one of the links. This result
is presented in the following theorem. 
\begin{theorem} \label{thm:wsum:1}
For  $0< \alpha < 1$, we have 
$C_{sum}(1,\alpha,\psi_1, \psi_2) = C_{sum}(1,1,\phi_1, \phi_2)$ 
where $\phi_1$ and $\phi_2$ are two derived CDFs given by
\begin{align}
\phi_1(h_1) &= \psi_1(h_1) \label{eq:phi1}\\ 
\phi_2(h_2) &= \alpha \psi_2 (h_2) + (1-\alpha),\, h_2 \geq 0. \label{eq:phi2}
\end{align}
\end{theorem}
Before we prove this result, a few remarks are in order. First of all, we already
know an optimal rate-allocation achieving the sum-capacity for any given set of
CDFs from the results of the previous section. 
Thus, evaluating the sum-capacity over $\phi_i, i=1,2$ is straightforward.
Second, only one of the CDFs need to be transformed to obtain the solution. The
transformation first scales the CDF and then shifts it appropriately 
to maintain its maximum height at unity, ensuring  a valid CDF after the transformation. 
This is illustrated in Fig.~\ref{fig:illus:2}, where $\phi_2$ is derived from $\psi_2$.


{\em Proof of the Achievability:}
Since an outage-free rate-allocation does not lead to outage in any fading block, it remains outage-free
even if we change the underlying fading distribution, provided the
respective supports of the distributions do not enlarge. Thus the optimal sum-capacity achieving rate-allocation
for $\phi_i, i=1,2$ is also an outage-free rate allocation under $\psi_i, i=1,2$, however this
may not be an optimal sum-capacity achieving rate allocation for the latter when $\alpha < 1$. 
Nevertheless, our interest is  in achieving the $(1,\alpha)$-weighted sum-capacity for $\psi_i, i=1,2$, 
and for that the optimal sum-capacity achieving rate allocations for $\phi_i, i=1,2$ suffice. 
These rate-allocations are given by
\begin{align} \label{eq:rate:cpreg:2}
R_i(h_i) = \int_0^{h_i} \dfrac{yP_i}{1+y^2P_i+(\phi_j^{-1}(\phi_i(y)))^2P_j}\,\mathrm{d}y,
\end{align}
where $i=1,2, j=1,2, i\neq j$, and $\phi_1(.),\phi_2(.)$ are as defined in (\ref{eq:phi1}) and (\ref{eq:phi2}).

%
%
%
%
Let us now show that $C_{sum}(1,\alpha,\psi_1,\psi_2) \geq C_{sum}(1,1,\phi_1,\phi_2)$.
%
Using the rate allocations~\eqref{eq:rate:cpreg:2}, albeit in channels $\psi_1$ and $\psi_2$, we get 
\begin{align*}
\mathbb E [R_1 + \alpha R_2]    
	&= \int_0^1 R_1(\psi_1^{-1}(x))dx + \alpha \int_0^1 R_2(\psi_2^{-1}(x)) dx \\
	&= \int_0^1 R_1 (\phi_1^{-1}(x)) dx + \alpha \int_0^1 R_2(\phi_2^{-1}(1-\alpha+\alpha x)) dx \\
	&= \int_0^1 R_1(\phi_1^{-1}(x))dx + \int_{1-\alpha}^1 R_2(\phi_2^{-1}(y))dy \\
	&= \int_0^1 R_1(\phi_1^{-1}(x))dx + \int_{0}^1 R_2(\phi_2^{-1}(x))dx \\
	&=  C_{sum}(1,1,\phi_1, \phi_2).
\end{align*}
where the second last equality followed from the fact that $\phi_2^{-1}(x)=0 $, for $x <  {1-\alpha}$.
Notice that $R_i(\cdot),i=1,2$ are chosen in ~\eqref{eq:rate:cpreg:2} as the 
sum-capacity achieving rate-allocation for the  CDFs $\phi_i,i=1,2$.
This completes the achievability proof.
\hfill \rule{2mm}{2mm} 

{\em Proof of the converse:}
We will now show that 
$$C_{sum}(1,\alpha,\psi_1,\psi_2) \leq C_{sum}(1,1,\phi_1,\phi_2).$$
Using the definitions in \eqref{eq:cdf:inv}
{\allowdisplaybreaks
\begin{align}
\mathbb{E}[R_1] +\alpha \mathbb{E}[R_2] 
&=\int_0^1 R_1(h_1(x)) dx\, +\alpha\int_{\bar\alpha}^1 R_2\left(h_2\left(\frac{y-\bar\alpha}{\alpha}\right)\right) \frac{dy}{\alpha}\notag \\
&=\int_0^{\bar\alpha} R_1(h_1(x)) dx+ \int_{\bar\alpha}^1 R_1(h_1(x)) dx\, 
		+\int_{\bar\alpha}^1 R_2\left(h_2\left(\frac{y-\bar\alpha}{\alpha}\right)
		\right) dy \label{eq:up:bnd:2}\\
&\leq \int_0^{\bar\alpha}\frac 12 \log(1+h_1(x)^2P_1)dx  \notag \\
	&\phantom{wwww}+\int_{\bar\alpha}^1 \frac 12 \log\left(1+h_1^2(x)P_1+h_2^2\left(\frac{x-\bar\alpha}{\alpha}\right)P_2\right).\label{cpregout} \\
&= C_{sum}(1,1,\phi_1, \phi_2).
\end{align}
}
In the above, the first step employed a simple coordinate scaling and translation, and 
the last step used  point-wise sum-rate bounds for Gaussian MAC, in particular by combining the last two integral terms in
\eqref{eq:up:bnd:2}, see Figure~\ref{fig:illus:2} for a visual verification. 
This completes the proof of the converse, and  thus Theorem~\ref{thm:wsum:1} is also proved.
\hfill \rule{2mm}{2mm}

\subsection{Numerical Example} \label{sec:sim:1}
\begin{figure}[htbp]
\begin{center} 
\includegraphics[scale=0.5]{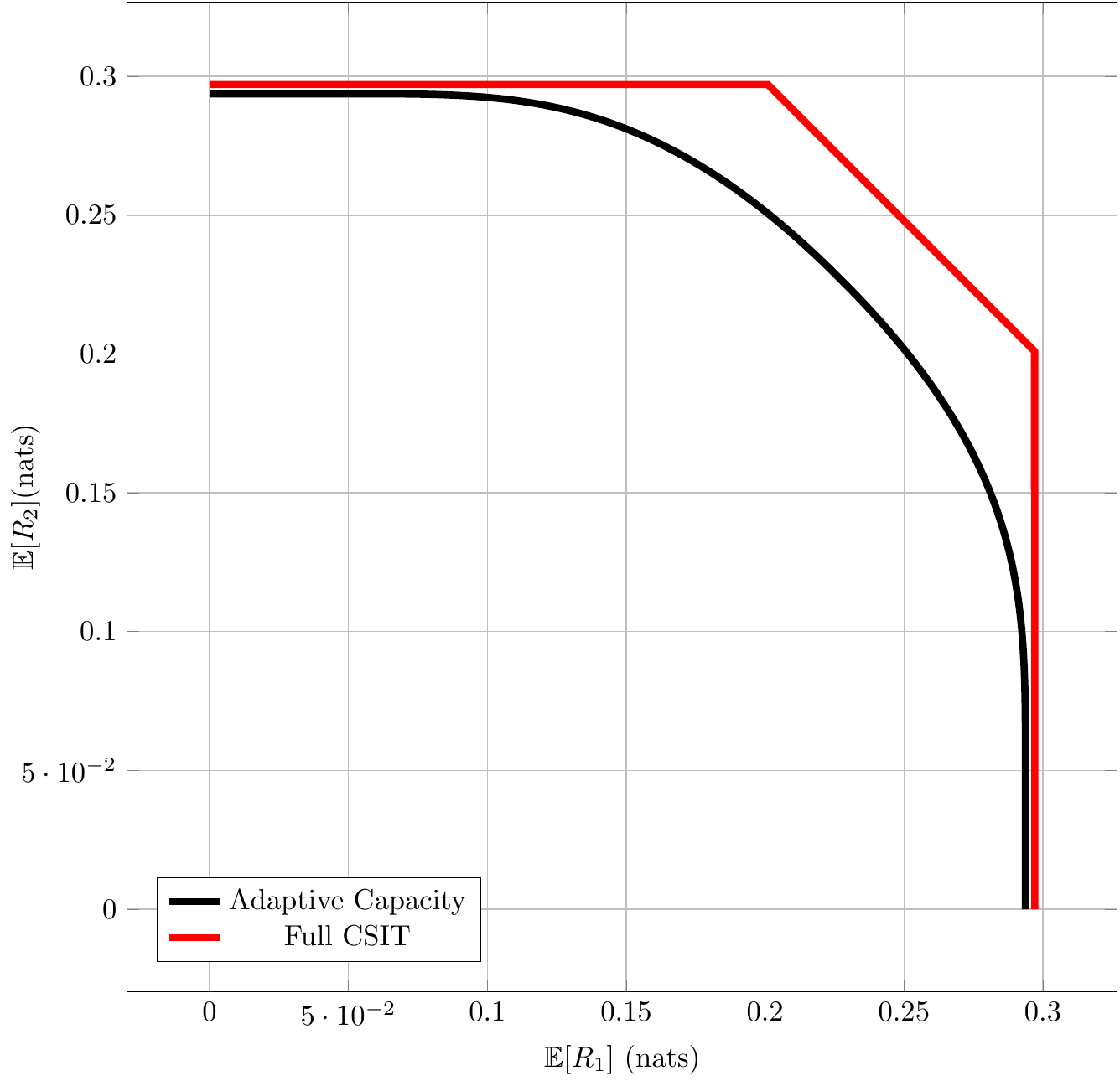}
\caption{ Adaptive capacity region when both the users have Rayleigh fading distributions with second moment $1$  \label{fig:capa:1}}
\end{center}
\end{figure}

It is of interest to
characterize the adaptive capacity region for some practical models. Consider a slow-fading
MAC with independent and identical Rayleigh distributed links. Figure~\ref{fig:capa:1} sketches
the capacity region for a transmit power $P_1=P_2=1$. The  variance of the
fading coefficient is taken to be $0.214$ (second moment = $1$).
Notice that the results known so far in literature were only successful in identifying a 
sum-capacity achieving rate-pair \cite{SDP13,GamalKim11},
whereas our current result obtain the full-capacity region.
For comparison, we also show the full CSI capacity region under no power-control. 
Note that even for maximizing the sum-rate, the full-CSI scheme is
different from the one where only the best user transmits \cite{TseHanly98} 
since we do not allow power control. 
The best scheme for full CSI can be numerically determined, we omit the details.

So far our results targeted a fixed transmit power. We will extend this in the next section to incorporate
power control by the transmitters.

\section{Power controlled Adaptive Capacity Region }
\label{sec:PCReg}

The adaptive sum-capacity of a fading Gaussian MAC with Individual 
CSI was described in section~\ref{sec:sum_capacity}, where
 for each user, the transmit power was fixed across blocks. However, it is
well known that a power control strategy which adapts the transmit powers
based on the fade values can significantly improve the
transmission rates for many systems, for example a MAC with
full CSIT~\cite{KnoppHumblet95}.  Similar improvements are also expected
in the distributed CSI MAC.
In this section, we allow power control, and compute the so called
\emph{power controlled adaptive capacity region} of a two user MAC
with individual CSI. The optimal power control law for the special case of
identical fading statistics across users were already derived in \cite{IPD12}.
Here we consider arbitrary but discrete fading statistics across the users.
The assumption of  discrete fading states is more of a technical requirement 
for the proof. 
Notice that even for  real-valued models, a power-rate strategy based
on  discretized fading states can closely match the actual performance.
We further restrict the exposition to a two user MAC, results for
many users follow along similar lines.

As a first step, we give a simple extension of our results in 
Sections~\ref{subsec:disc:two}  and \ref{subsec:dis:mult} to incorporate any given
set of power control laws at the users. Assume that for $i=1,2$,
user~$i$ employs a power allocation function $P_i(h_i)$ which
also meets the long-term average power constraint $P^{avg}_i$.
Let $C_{sum,P_1,P_2} (1,\alpha, \psi_1, \psi_2)$ denote 
the adaptive weighted sum-capacity under the given pair of power allocation
functions $P_i(h_i), i=1,2$ at the respective transmitters. 
Using this notation, for fixed transmit powers (as in Section~\ref{subsec:disc:two}) 
we will write $C_{sum,c_1,c_2} (1,\alpha, \psi_1, \psi_2) $ 
where $c_i$ is the power employed by user~$i$ across fading states. 

The quantity $C_{sum,P_1,P_2} (1,\alpha, \psi_1, \psi_2)$ can be
evaluated as follows.  Let us define
$g_i = \sqrt{{h_i^2 P_i(h_i)}}$ and consider a new block fading MAC
with fading vector $\mathbf g$ and fixed transmit powers of unity across
all fading realizations, i.e.  there is no power adaptation in this new
MAC model.  For such a fixed transmit power system, 
we already know the weighted sum-rate
from Theorem~\ref{thm:wsum:1}. The following lemma is immediate from
this discussion.

%
\begin{lemma} \label{lem:pow:ctrl}
$$
C_{sum,P_1,P_2} (1, \alpha, \psi_1, \psi_2) = C_{sum,1,1} (1, 1, \phi_1, \phi_2)  
$$
with
\begin{align}
\phi_1(g_1) = \nu_1(g_1) \,;\,\,
\phi_2(g_2) = \alpha \nu_2 (g_2) + (1-\alpha),\, g_2 \geq 0,
\end{align}
and $\nu_i(g_i)$ is the CDF of $\sqrt{{H_i^2 P_i(H_i)}}$ for $i \in\{1,2\}$.
\end{lemma}
%
%
%
While the above lemma is simple, it is extremely useful  in the  sense 
of separating the power-control and rate maximization.
In particular, our results in Section~\ref{sec:adap:capa:reg} can easily specify the
optimal rate-adaptation for any given set of power-control laws. The question
now is about optimizing the power-allocation. Unfortunately we do not have a
closed form solution for optimal power-control, except for identical statistics 
across users~\cite{IPD12}. Therefore, one needs to resort to numerical techniques
to evaluate the optimal laws. This may appear formidable due to the polymatroidal constraints 
imposed on the possible rate-choices. An alternate way is to  identify some thumb-rule for
power-allocation (see~\cite{ShamaiTelatar99}) and then choose the optimal
rate-adaptation. Iterative techniques based on gradient based search is 
a widely used technique to identify optimal laws.  

Once the power control laws are given, the proposed optimal scheme 
does rate-allocation in the increasing order of  $h_i^2P_i(h_i)$. 
If the order among $h_i^2P_i(h_i)$ is preserved while any 
algorithm searches for an optimal power control
then searching time and effort can be considerably reduced.
The reason for this can be explained better for the evaluation of sum-rate.
Notice that while evaluating the sum-rate, changing the power-allocation
for a particular fading state will have a localized effect on the sum-rate.
This can be visualized with the aid of  Figure~\ref{fig:rate:assign}.
Suppose the power allocation of any state of user~$1$ is changed in such
a way that the horizontal cuts of the CDF still stay the same. Then the
sum-rate of those states of user~$2$ which shares a horizontal cut
on the respective CDFs will be affected by the new rate-allocation. 
All other sum-rate values stay the same. Similar arguments apply when a power
allocation is changed for a pair of states, while preserving the average transmit power. 
This allows the numerical solutions to proceed by localized searches, 
a very powerful advantage in locating the optimal power allocation.
The complexity of the search becomes of the order of $|\Gamma|$, which
is the number of distinct state-pairs intersected by the horizontal cuts on 
the CDFs, this is  evident from the right most diagram in Figure~\ref{fig:rate:assign}.
From the standpoint of  preserving the orders of the received powers, 
the following theorem is important, 
as it guarantees  the existence of an optimal power allocation  with the desired 
monotonicity property.



\begin{lemma} \label{lem:monpwr}
There exists an optimal power-allocation in which
$h_{i}^2P_i(h_{i})$ is a non-decreasing function of $h_{i}$ for $i=1,2$.  
\end{lemma}
\begin{IEEEproof}
Let there be $k_i$ values for $H_i$ with probabilities $p_{ij}, 0 \leq j \leq k_i -1$.
Let $P_1(\cdot),$ and $P_2(\cdot)$ be two power allocation
functions at the respective users. Denote  $H_i^2 P_i(H_i)$ by $G_i$.

In order to avoid the notations from blowing up, we
assume that  $G_i$ has its mass on $k_i$ distinct values, say $\{u_{ij}, 0 \leq j \leq k_i -1\}$ 
ordered in the ascending fashion.
The assumption of distinct values is simply to create a bijection from $H_i$ to $G_i$,
the exposition becomes simpler by this. Nevertheless, the proof applies more
generally, with some renaming of the indices.

We proceed by contradiction.  Consider two fade values  $h_1^{\prime}, h_1^{\prime\prime}$ such that
$g_1^{\prime}=h_1^{\prime 2}P_1(h^{\prime}_1) > g_1^{\prime\prime }=h_1^{\prime\prime 2} P_1(h^{\prime \prime}_1)$, 
though $h^{\prime}_1 < h^{\prime\prime}_1$. 
Assume that $g_1^{\prime}$ and $g_1^{\prime\prime}$ are the
first pair of adjacent values to have this property, i.e. there is no value for $h_1^2P_1(h)$ in the open
interval $(g_1^{\prime\prime}, g_1^{\prime})$. There is no generality lost here, 
if there is a value in the middle, we can redefine
$g^{\prime}_1$ or $g_1^{\prime \prime}$ appropriately and choose the corresponding $h_1$ values
to pair with it. 

 Let the probabilities of $g_1^{\prime}$ and $g_1^{\prime\prime}$ be
 $p^{\prime}$ and  $p^{\prime\prime}$ respectively, which are assumed to be positive. 
%
Denote the probability mass function of $G_i$ by $Q_i(\cdot)$. We will scale and
shift the CDF of $G_2$ to take care of the weighted averages, as in Section~\ref{sec:adap:capa:reg}.
For the $(1,\alpha)$ weighted sum-rate with $\alpha\in[0,1]$, 
let us define $q_{20} := 1-\alpha  + \alpha \,Q_2(u_{20}) \indicator{u_{20}=0}$, and for $1\leq j \leq k_2-1$,
\begin{align}
q_{2j} = \begin{cases} \alpha Q_2(u_{2(j)}) \text{ if } u_{20} = 0 \\
	\alpha Q_2(u_{2(j-1)}) \text{ otherwise}.
	\end{cases}
\end{align}
The second operation above shifts the mass function to the right so as to accommodate a mass at zero,
 if this is not already in the
support. In this case, we should also define $q_{2k_2}$, but since
$q_{2k_2} = 1- \sum_{j < k_2} q_{2j}$, a new definition will turn out 
redundant in our rate-allocation scheme. 
Let $q_{1j} := Q_1(u_{1j}), 0 \leq j \leq k_1-1$. Similar to \eqref{eq:def:alpha} -- \eqref{eq:def:beta}, 
we can define
\begin{align}
 \alpha_k &= \sum_{j=0}^k q_{1j}\,,\,\, 0 \leq k \leq k_1-1, \\
\beta_k &= \sum_{j=0}^k q_{2j} \,,\,\, 0 \leq k \leq k_2-1, 
\end{align}
and let $\Gamma= \{\gamma_i| 0 \leq i \leq |\Gamma|-1 \} := \{\alpha_i| 0\leq i \leq k_1-1\}
\cup \{\beta_i| 0 \leq i \leq k_2-1\}$ be an ordered set with the elements following an ascending
order.


Let $\nu_i, i=1,2$ denote the CDFs of the respective mass functions $\{q_{ij},1 \leq j \leq k_i-1\}$.
Now for $i=1,2$, let us define the inverse CDF values for user~$i$  as
\begin{align} \label{eq:disc:cdfrind}
g_{ij} & = \sup\{g| \nu_i(g) < \gamma_j\}.
\end{align}
This definition implies that
\begin{align}
 \sum_{k:g_{ik}=h_{ik}^2P_i(h_{ik})}[\gamma_{k+1}-\gamma_k]=q_{ik}.
\end{align}
Since we are interested in the weighted sum-capacity, let us assume an optimal rate-allocation 
according to Theorem~\ref{thm:wsum:1}. Equivalently, by Lemma~\ref{lem:pow:ctrl}, we can use the allocations 
\eqref{eq:th1a}--\eqref{eq:th1c} for  the CDFs $\nu_1,\nu_2$ and unit power at the transmitters.
%
Thus, the optimum weighted sum-rate with the power allocation $P_1(\cdot), P_2(\cdot)$ is given by
\begin{align}
\eE(R_1)+ \alpha \eE(R_2)
=& \sum_{k=0}^{|\Gamma|-1} \left( R_1(g_{1k})  + R_2(g_{2k})\right) [\gamma_{k+1} - \gamma_k] \label{eq:allsum} \\
=&\sum_{k\in A^{\prime\prime}} \frac 12 \log\left(1+g_1^{\prime\prime }+g_{2k}\right)[\gamma_{k+1}-\gamma_k]  
	+\sum_{k \in A^{\prime}} \frac 12 \log\left(1+g_1^{\prime}+g_{2k}\right)[\gamma_{k+1}-\gamma_k]  + R_r \notag \\
=& \sum_{k\in A^{\prime\prime}} \frac 12 \log\left(1+h_{1}^{\prime\prime 2}P_1(h_{1}^{\prime\prime})+g_{2k}\right)[\gamma_{k+1}-\gamma_k]  +\sum_{k \in A^{\prime}} \frac 12 \log\left(1+h_{1}^{\prime 2}P_1(h_{1}^\prime) +g_{2k}\right)[\gamma_{k+1}-\gamma_k]  + R_r \label{eqn3old},
\end{align}
where $A^{\prime}:=\{k:g_{1k}=h_1^{\prime 2} P_1(h^{\prime}_1)\}$, $A^{\prime\prime}:=\{k:g_{1k}=h_1^{\prime\prime 2} P_1(h^{\prime\prime}_1)\}$, 
and $R_r$ denotes the rest of the summation in \eqref{eq:allsum}.
Let  $P_{1\epsilon}(.)$ be a new power allocation such that
\begin{align}
P_{1\epsilon}(h_1^{\prime\prime})=P_1(h_1^{\prime\prime})+\epsilon; \, P_{1\epsilon}(h_1^{\prime})=P_{1}(h_1^{\prime})-\frac{\epsilon p^{\prime\prime}}{p^{\prime}}; \,
P_{1\epsilon}(h_1)=P_{1}(h_{1}), \text{ for } h_1 \neq h_1^{\prime}, h_1^{\prime\prime}.\notag
\end{align}
It is easy to see that this new power allocation satisfies the power constraints.
Here $\epsilon>0$ is chosen small enough such that $h_1^2P_1(h_1)$ and $h_1^2P_{1\epsilon} (h_1)$ 
occupy the same place in the ordered list of received powers of user~$1$.
In particular,
the horizontal levels $\gamma_k$ in these two CDFs are the same and the sets $A^{\prime}$ and $A^{\prime\prime}$ are also the same.
Hence, the optimum weighted sum-rate with the new power allocation $P_{1\epsilon}(\cdot)$ is given by
\begin{align}
S(\epsilon):= &\eE(R_1)+\alpha \eE(R_2)\notag\\
=& \sum_{k\in A^{\prime\prime}} \frac 12 \log\left(1+h_{1}^{\prime\prime 2}P_1(h_{1}^{\prime\prime})+\epsilon h^{\prime\prime 2}+g_{2k}\right)[\gamma_{k+1}-\gamma_k] \notag  \\ 
 &\phantom{wwwww}	+\sum_{k \in A^{\prime}} \frac 12 \log\left(1+h_{1}^{\prime 2}P_1(h_{1}^\prime)-\frac{\epsilon p^{\prime\prime}}{p^{\prime}} h_1^{\prime 2} +g_{2k}\right)[\gamma_{k+1}-\gamma_k] 
 + R_r \label{eqn3}.
\end{align}
Taking derivative, we have
\begin{align}
 S^{\prime}(\epsilon)= & \sum_{k\in A^{\prime\prime}}\frac{h_1^{\prime\prime 2}}{2\left(1+h_{1}^{\prime\prime 2}P_1(h_{1}^{\prime\prime})
	+\epsilon h^{\prime\prime 2}+g_{2k}\right)}[\gamma_{k+1}-\gamma_k] \notag\\
&\phantom{wwwww}-\sum_{k\in A^{\prime}} \frac{p^{\prime\prime}}{p^{\prime}} \frac{h_{1}^{\prime 2}}{2\left(1+h_{1}^{\prime 2}P_1(h_{1}^\prime)-\frac{\epsilon p^{\prime\prime}}{p^{\prime}} h_1^{\prime 2} +g_{2k}\right)}[\gamma_{k+1}-\gamma_k]. \label{derepsilon}
\end{align}
At $\epsilon=0$, 
\begin{align}
 S^{\prime}(0)=&\sum_{k\in A^{\prime\prime}} \frac{h_1^{\prime\prime 2}}{2\left(1+h_{1}^{\prime\prime 2}P_1(h_{1}^{\prime\prime})+g_{2k}\right)}[\gamma_{k+1}-\gamma_k] \notag\\
&\phantom{wwwww}-\sum_{k\in A^{\prime}} \frac{p^{\prime\prime}}{p^{\prime}} \frac{h_{1}^{\prime 2}}{2\left(1+h_{1}^{\prime 2}P_1(h_{1}^\prime) +g_{2k}\right)}[\gamma_{k+1}-\gamma_k]\notag.
\end{align}
Since $h_1^{\prime 2} P_1(h_1^\prime) > 
h_1^{\prime\prime 2} P_1(h_1^{\prime\prime})$, and our optimum rate-allocation
algorithm assigns rates in increasing order of the received power for
both the users, $k^{\prime} > k^{\prime\prime}$ for any $k^{\prime}\in A^{\prime}, k^{\prime\prime} \in A^{\prime\prime}$. Thus, for such
$k^{\prime}, k^{\prime\prime}$, we have $g_{2k^{\prime}} \geq g_{2k^{\prime\prime}}$.
So, if $\bar{k}:=\max A^{\prime\prime}$, then
\begin{align*}
 g_{2k^{\prime}} \geq g_{2\bar{k}} \text{ for } k^{\prime} \in A^{\prime} \text{ and } g_{2k^{\prime\prime}}\leq g_{2\bar{k}} \text{ for } k^{\prime\prime} \in A^{\prime\prime}.
\end{align*}
Thus, we have
\begin{align*}
 S^{\prime}(0)\geq & \sum_{k:g_{1k}=g_1^{\prime\prime}} \frac{h_1^{\prime\prime 2}}{2\left(1+h_{1}^{\prime\prime 2}P_1(h_{1}^{\prime\prime})+g_{2\bar{k}}\right)}[\gamma_{k+1}-\gamma_k] \\
&\phantom{wwwww}-\sum_{k:g_{1k}=g_1^\prime} \frac{p^{\prime\prime}}{p^{\prime}} \frac{h_{1}^{\prime 2}}{2\left(1+h_{1}^{\prime 2}P_1(h_{1}^\prime) +g_{2\bar{k}}\right)}[\gamma_{k+1}-\gamma_k].
\end{align*}
It follows from the definition of $\gamma_k $ that,
\begin{align} \label{eq:gamma:diff}
 \sum_{k\in A^{\prime}}[\gamma_{k+1}-\gamma_k]=p^{\prime}  \text{ and } 
\sum_{k\in A^{\prime\prime}}[\gamma_{k+1}-\gamma_k]=p^{\prime\prime}.
\end{align}
 Hence,
\begin{align}
 &S^{\prime}(0)\geq \frac{p^{\prime\prime}h_1^{\prime\prime 2}}{2\left(1+h_{1}^{\prime\prime 2}P_1(h_{1}^{\prime\prime})+g_{2\bar{k}}\right)}-\frac{p^{\prime\prime}h_{1}^{\prime 2}}{2\left(1+h_{1}^{\prime 2}P_1(h_{1}^\prime) +g_{2\bar{k}}\right)}\notag.
\end{align}
Since $h_1^{\prime 2} P_1(h_1^\prime) > h_1^{\prime\prime 2} P_1(h_1^{\prime\prime})$, and $h_1^{\prime\prime} > h_1^{\prime}$, we conclude that
$S^{\prime}(0) > 0$.

Note that $S^{\prime}(\epsilon)$ defined in \eqref{derepsilon} is a continuous, monotonically decreasing  and differentiable 
function for $\epsilon \geq 0$. 
This shows that in an optimum power allocation, $P_1(\cdot)$ is a
non-decreasing function. 

We should now show that the second user's power allocation leads to  a non-decreasing $h^2P_2(h)$. This can be obtained by
similar arguments as above, the main change is that the probabilities for non-zero values of $G_2$ have to be 
scaled by $\alpha$ in the computations. In particular, the values $p^{\prime}$ and $p^{\prime\prime}$ 
in \eqref{eq:gamma:diff} will be scaled by $\alpha$, without affecting the overall sign of the quantities.
We do not repeat all the arguments here. This completes the proof of Lemma~\ref{lem:monpwr}.  
\end{IEEEproof}
The advantage of Lemma~\ref{lem:monpwr} can be gleaned by considering the sum-rate evaluation.
The lemma implies that under any optimal power allocation, $h_{ik}^2P_i(h_{ik})$ is increasing. 
This ensures that the horizontal levels $\gamma_i$ of the CDF of $G_i$ 
remain fixed for any optimal power allocation for sum-rate, 
and these are same as the horizontal levels in the CDF $\psi_i$ (see Remark~\ref{rem:hor:cut}). 
Hence, the sum-rate expression defined with these $\gamma_i$  in \eqref{eq:allsum} and $\alpha=1$ 
is a valid objective function  for maximization over the set of all possible  power 
allocations which are candidates for optimality. 
The constraint set is defined by the  average power constraints and the conditions 
\begin{align}
 h_{ik}^2P_i(h_{ik}) \geq h_{ij}^2P_i(h_{ij}) \text{ if } h_{ik} \geq h_{ij},~k,j \in \{1,2,\cdots, n_i\}, ~i=1,2 \notag.
\end{align}
 It is easy to see that the objective function  is a concave function of the power variables and 
the constraints
 are  linear (hence convex) in the power variables $P_i(.)$. Hence, standard results in non-linear 
programming can be used to guarantee the convergence of 
a gradient based search algorithm for finding the power controlled adaptive sum-capacity, 
in which the power is modified in each step  of the  iteration  depending on the direction of the 
gradient of the objective function. These algorithmic aspects are outside the purview of the current paper. However,
for illustration, we compute the power controlled capacity region of a two-state fading model, which is an example
studied in  \cite{GamalKim11}.  
\begin{figure}[htbp]
\begin{center}
\includegraphics[scale=0.9]{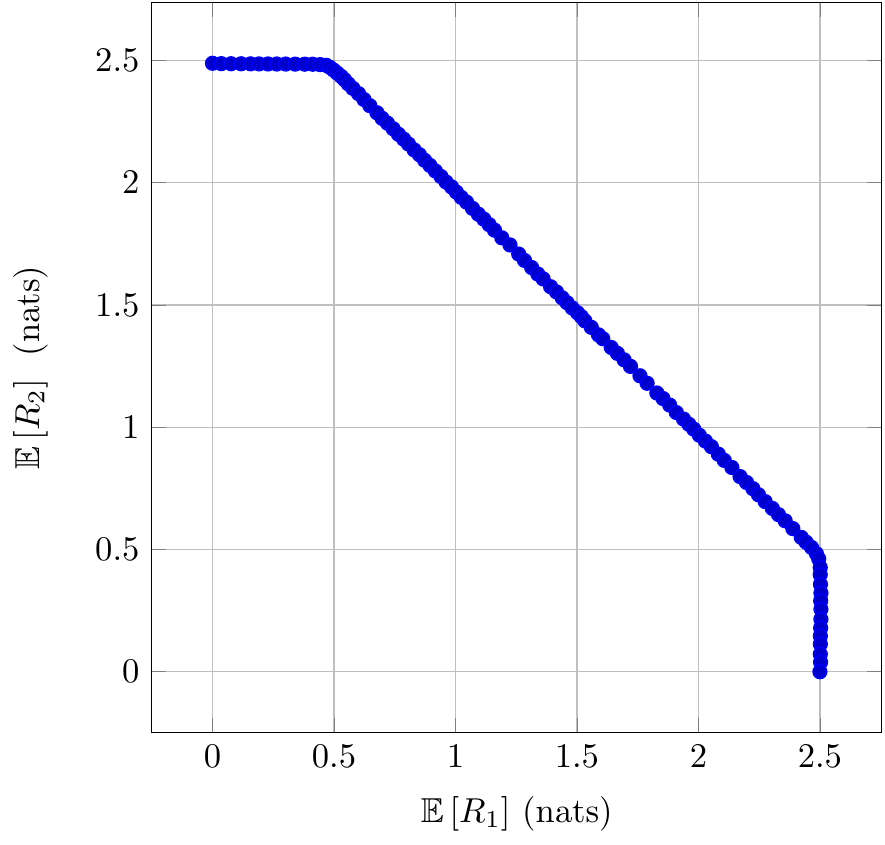}
\end{center}
\caption{Power-Adaptive capacity region, Users identical with two fading states  $h_b=1$, $h_g=2$,  $P(h_b)=0.2$  and $P_{avg}=10$}
\label{fig:cpregdis}
\end{figure}
Notice that, the same procedure can easily identify the capacity region for several discrete models.  
While the numerical study in \cite{GamalKim11} only targets the
sum-capacity for a two-state model, the full power-controlled adaptive  capacity-region for 
channels with several fading states  can be computed by the techniques presented here. 

\begin{figure}[htbp] 
\centering\includegraphics[scale=0.8]{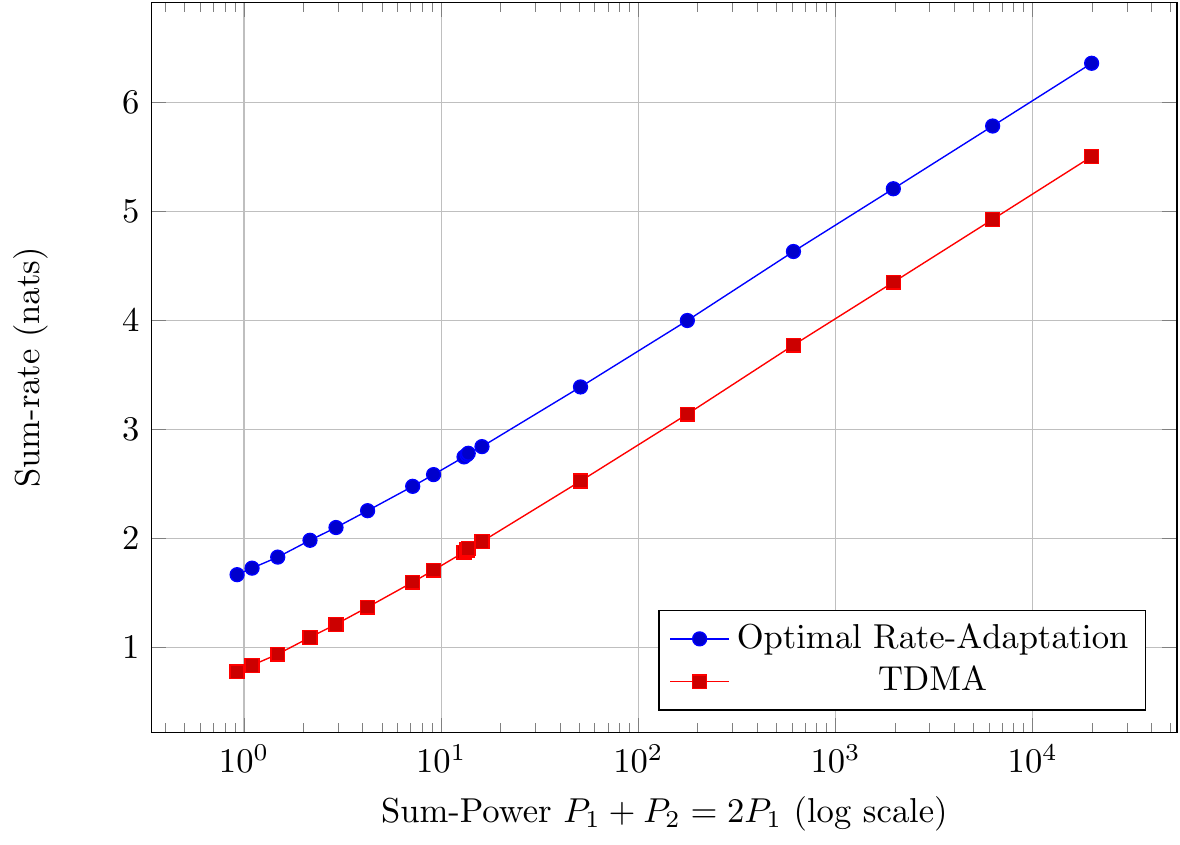}
\caption{Sum-rate for
$\psi_2$ Rayleigh with second moment $10$, and $\psi_2$ uniform in $[0,\sqrt 3]$\label{fig:tdm:pow:ctrl}}
\end{figure}

In the remaining of the section, let us compare the optimal power-controlled 
TDMA  with the proposed schemes here. 
Notice that the optimal TDMA power control can be
sub-optimal when used in conjunction with other rate-allocation mechanisms, 
however it still serves as a benchmark for performance comparison. 
In particular, given a  time-sharing parameter, the optimal TDMA power control follows a single user water-filling structure,
 with  appropriate water-levels chosen to respect the average power constraints at the users.
To illustrate the performance, let us 
consider a Rayleigh fading link with second moment of $10$, and another link uniformly distributed
in $[0,\sqrt{3}]$. Under equal power constraints, the optimal sum-rate for generalized TDMA is
plotted in Figure~\ref{fig:tdm:pow:ctrl} against the sum-power. Now, for the same 
power-control law, we can use the rate-adaptation mechanism given by Lemma~\ref{lem:pow:ctrl}.
This is easily achieved by defining the $\sqrt{P_i(H_i)}H_i$ as the new fading coefficient,
where $P_i(H_i)$ is the optimal TDMA water-filling power-control function. 
It is clear from Figure~~\ref{fig:tdm:pow:ctrl}
that the schemes proposed here  outperform the best TDMA schemes. Furthermore, employing the best
power control schemes can make the rates even better, showing the suboptimality of TDMA
in such distributed settings.

%




\section{Additional CSI on the Other Links} 
\label{sec:extra:csi}
 In this section, we assume that each transmitter also has
some partial CSI  of the other links, in addition to the complete knowledge
of its own link. Let us again consider a two user block fading MAC for simplicity. 
To start with, we also assume that the
additional partial CSI from the other link is generated by a quantizer.



Let $\hat{h}_1$ denote the quantized value of $h_1$ which is known to user~$2$,
and similarly $\hat{h}_2$ as  the quantized value of $h_2$ available at 
user~$1$. Consider 
a pair of power allocation functions $P_1(h_1,\hat h_2)$ and 
$P_2(\hat h_1, h_2)$ for the users 1 and 2 respectively.
As in the last section, let $\psi_i(h_i)$ denote the fading CDF of the 
$i^{th}$ user and $\psi(h_1,h_2)=\psi_1(h_1)\psi_2(h_2)$  denote their joint CDF 
(i.e. independently fading links).

Imagine that the values of $h_1$ are partitioned into $B_1$ non-overlapping sets 
$S_1, \cdots, S_{B_1}$ which are mapped to different output values by
the quantizer ($\hat h_1$).
We denote the minimum fading magnitude in the set $S_i$ by $m_j$. 
Similarly, let $T_1, \cdots, T_{B_2}$ represent the $B_2$ partitions of $h_2$, 
and $n_k=\min T_k$.
We define $q_{1i} := Pr (H_1 \in S_i)$ for $1\leq i\leq B_1$ and $q_{2j} := Pr (H_2 \in T_j)$ for $1\leq j\leq B_2$.
We can now write,
\begin{align}
 \eE(R_1(H_1,\hat H_2)+ \alpha R_2(\hat H_1,H_2))
&=\int  \int  R_1(h_1,\hat h_2) d{\psi(h_1,\hat h_2)}  
		+\alpha \int   \int R_2(\hat h_1, h_2) \,d{\psi(\hat h_1, h_2)}\notag\\
&=\sum_{i,j}\int_{S_i}  \int_{T_j}  R_1(h_1,\hat h_2)\,d{\psi(h_1,\hat h_2)} 
		+ \alpha \sum_{i,j}\int_{S_i}  \int_{T_j}R_2(\hat h_1,h_2)\,d{\psi(\hat h_1,h_2)}\notag\\
&=\sum_{i,j}\bigg[q_{2j}\int_{S_i} R_1(h_1,n_j)\,d{\psi_1(h_1)} 
	+ \alpha q_{1i}\int_{T_j} R_2(m_i,h_2)\,d{\psi_2(h_2)}\bigg] \notag\\
&=\sum_{i,j} q_{1i}q_{2j} R_{sum}^{(i,j)}(1,\alpha) \notag
\end{align}
where
\begin{align}
R_{sum}^{(i,j)}(1,\alpha) = & \int_{S_i} R_1(h_1,n_j)\,\frac{d{\psi_1(h_1)}}{q_{1i}} 
		+ \alpha \int_{T_j} R_2(m_i,h_2)\,\frac{d{\psi_2(h_2)}}{q_{2j}} \label{eq:comb:pcsi}
\end{align}
is the weighted sum-rate under the condition $H_1 \in S_i, H_2 \in T_j$.
Note that both the users know the values of $i$ and $j$. Thus for different
values of $(i,j)$, the pairs of functions $(R_1(\cdot,n_j), R_2(m_i,\cdot))$ 
can be optimized independently. 
Notice that 
$$
\int_{T_j} d{\psi_2(h_2)} = q_{2j} \text{ and } \int_{S_i} d{\psi_1(h_1)} = q_{1i}.
$$
So each integral in \eqref{eq:comb:pcsi} is evaluated with respect 
to a conditional distribution.  Hence \eqref{eq:comb:pcsi} is of the same 
form as \eqref{eq:wsum:1}, and for each $i,j$, the expression in \eqref{eq:comb:pcsi}
can be maximized using Theorem~\ref{thm:wsum:1}, this will in turn maximize the
overall weighted sum-rate.
Let us  demonstrate the utility of additional CSI by numerical comparisons.  

\subsection{Numerical Example} \label{sec:sim:2} 
\begin{figure}[htbp]
\begin{center}
\includegraphics[scale=0.7]{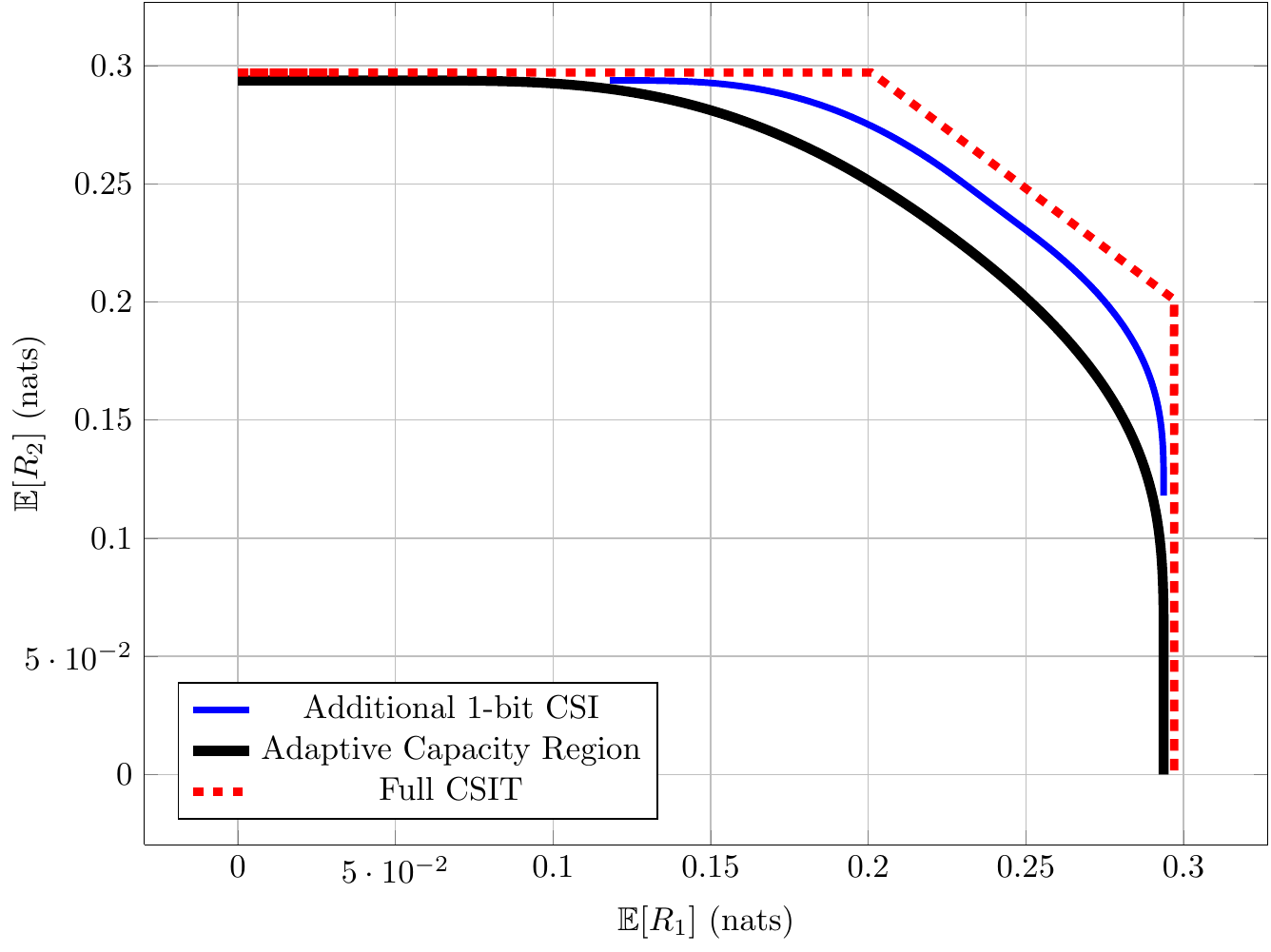}
\caption{Capacity Enlargement by Additional CSI\label{fig:two}}
\end{center}
\end{figure}

In this subsection, we consider the same example setup in Section~\ref{sec:sim:1},
however $1$~bit of partial CSI from the other link is additionally made available at each transmitter. 
The single bit is obtained
by comparing the CSI against a known threshold. Figure~\ref{fig:two} compares the enlargement of the
adaptive capacity region with $1$~bit additional partial CSI. The threshold for the quantizer was arbitrarily taken to
be $0.4$ for demonstration purpose. In an application where the threshold
can be chosen by the designer/users, the best choice of this threshold is
an important question that deserves further investigation.

Under any given power-allocation schemes, we can write each integral
in terms of the received powers and use the result 
described in Section~\ref{sec:PCReg} to maximize the weighted sum-rate.
Thus the  adaptive capacity region under additional partial CSI can be
computed in an efficient manner. 
Extensions to multiple users and 
other models where the additional CSI is obtained by deterministic functions
of the fading coefficients etc follow along similar principles.
At the extreme case, where one user knows both the channels and other knows only
its own, the model becomes an asymmetric CSIT MAC \cite{CemalSteinberg05}.
The notions of adaptive and ergodic capacity coincides here and our techniques
can numerically solve the capacity region for this case. Let us also mention
about power control and additional CSI.

For discrete fading states  we can also extend our discussion to the power controlled adaptive
capacity.
For any fixed  power allocation $P_1(h_1,\hat{h_2})$ and $P_2(\hat{h_1},h_2)$ 
for the users~$1$ and~$2$ respectively, the adaptive sum capacity with partial CSI 
can be obtained using the distribution on received powers, as explained above. 
By applying Lemma~\ref{lem:monpwr}, an optimal power allocation can be shown to be 
such that, for fixed $\hat{h_2}$ , $h_1^2P_1(h_1,\hat{h_2})$ is  monotone 
increasing in $h_1$ in each of the regions $s_k$ . Likewise, for 
fixed $\hat{h_1}$, $h_2^2P_2(\hat{h_1},h_2)$ is monotone increasing in $h_2$ in each of the regions $t_k$.
 This leads to an expression for the expected sum rate involving the 
powers $P_1(h_1,\hat{h_2})$ and $P_2(\hat{h_1},h_2)$ as the variables, similar to the individual CSI case. The expression is a concave function of these power variables
and the power constraints are also linear (hence convex). 
 The power allocation which maximizes this expression  can be found by any of the methods for solving convex optimization problems, thus giving an algorithm for finding the power controlled adaptive sum capacity with partial CSI about the other link.

 \section{Conclusion}\label{sec:conclusion}
In this work,  we presented the adaptive capacity region of fading MACs with arbitrary
fading statistics for varying amounts of channel state information available at the transmitters.
The techniques also provided  the power-controlled adaptive capacity region for channels with
discrete states.
 
For the case of individual CSI,
the solution for the adaptive sum capacity (without power control) was presented in an elegant closed form for 
continuous valued fading distributions, and as an iterative rate allocation expression for
discrete fading states.  These formulas work for any number of users and link statistics.
Finding the adaptive capacity region with individual CSI amounts to finding the 
rate allocations that maximize the expected  weighted sum rate of the users.
 We have presented an outage free rate allocation strategy which can achieve any point on the boundary of the  
adaptive capacity region. Thus we have characterized the entire adaptive capacity region.

We also presented a 
result which reduces the problem of finding the power controlled 
adaptive weighted sum-capacity to a much simpler convex optimization problem with linear constraints.
The power controlled adaptive capacity when transmitters have additional partial 
CSI about the other links was discussed in section \ref{sec:extra:csi}.
While it is of interest to characterize the adaptive capacity region when the individual channel
knowledge is also not perfect, this can be handled by our techniques in several interesting cases.
In particular, any scenario where a user has more information than others about its link, in the sense that 
user~$i$ has
access to the information others have about $h_i$, can be covered by a suitable extension 
of our rate-allocation technique. This was not included in the current paper
due to the overwhelming amount of notations required, and also to keep the length of the submission
under control. The proposed techniques also enable the computation of 
 power controlled adaptive capacity region with varying amounts of transmitter CSI.
While we have focused on the safe mode of operation in this paper, 
it would be interesting to evaluate the power controlled adaptive capacity region   when
 outage is permitted for some  users and state-tuples, 
and this is a direction of research that we will pursue further.

\bibliographystyle{IEEEtran}

\begin{thebibliography}{10}
\providecommand{\url}[1]{#1}
\csname url@samestyle\endcsname
\providecommand{\newblock}{\relax}
\providecommand{\bibinfo}[2]{#2}
\providecommand{\BIBentrySTDinterwordspacing}{\spaceskip=0pt\relax}
\providecommand{\BIBentryALTinterwordstretchfactor}{4}
\providecommand{\BIBentryALTinterwordspacing}{\spaceskip=\fontdimen2\font plus
\BIBentryALTinterwordstretchfactor\fontdimen3\font minus
  \fontdimen4\font\relax}
\providecommand{\BIBforeignlanguage}[2]{{%
\expandafter\ifx\csname l@#1\endcsname\relax
\typeout{** WARNING: IEEEtran.bst: No hyphenation pattern has been}%
\typeout{** loaded for the language `#1'. Using the pattern for}%
\typeout{** the default language instead.}%
\else
\language=\csname l@#1\endcsname
\fi
#2}}
\providecommand{\BIBdecl}{\relax}
\BIBdecl

\bibitem{TseHanly98}
D.~Tse and S.~Hanly, ``Multiaccess fading channels. i. polymatroid structure,
  optimal resource allocation and throughput capacities,'' \emph{Information
  Theory, IEEE Trans on}, vol.~44, no.~7, pp. 2796 --2815, Nov. 1998.

\bibitem{TseViswanath05}
D.~Tse and P.~Viswanath, \emph{Fundamentals of Wireless Communication}.\hskip
  1em plus 0.5em minus 0.4em\relax Cambridge University Press, 2005.

\bibitem{ShamaiTelatar99}
S.~Shamai and E.~Telatar, ``Some information theoretic aspects of decentralized
  power control in multiple access fading channels,'' in \emph{Information
  Theory and Networking Workshop, 1999 IEEE}, June 1999.

\bibitem{DasNarayan02}
A.~Das and P.~Narayan, ``Capacities of time-varying multiple-access channels
  with side information,'' \emph{Information Theory, IEEE Transactions on},
  vol.~48, no.~1, pp. 4 --25, Jan. 2002.

\bibitem{KeStMe08}
G.~Keshet, Y.~Steinberg, and N.~Merhav, \emph{Channel Coding in the Presence of
  Side Information}.\hskip 1em plus 0.5em minus 0.4em\relax Now Publishers,
  Foundations and Trends® in Communications and Information Theory, 2007,
  vol.~4, no.~6.

\bibitem{LaSt13}
A.~Lapidoth and Y.~Steinberg, ``The multiple access channel with causal side
  information: double state,'' \emph{Information Theory, IEEE Transactions on},
  vol.~59, no.~1, pp. 32--50, 2013.

\bibitem{CemalSteinberg05}
Y.~Cemal and Y.~Steinberg, ``The multiple-access channel with partial state
  information at the encoders,'' \emph{Information Theory, IEEE Transactions
  on}, vol.~51, no.~11, pp. 3992 -- 4003, nov. 2005.

\bibitem{Jafar06}
S.~A. Jafar, ``Channel capacity with causal and non-causal state information- a
  unified view,'' \emph{Information Theory, IEEE Transactions on}, vol.~52,
  no.~12, pp. 5468--5474, 2006.

\bibitem{ZaPiSh13}
A.~Zaidi, P.~Piantanida, and S.~Shamai~(Shitz), ``Capacity region of
  cooperative multiple access channel with states,'' \emph{Information Theory,
  IEEE Transactions on}, vol.~59, no.~10, pp. 6153--6174, 2013.

\bibitem{BiPrSh98}
E.~Biglieri, J.~Proakis, and S.~Shamai, ``Fading channels:
  information-theoretic and communications aspects,'' \emph{Information Theory,
  IEEE Transactions on}, vol.~44, no.~6, pp. 2619--2692, Oct 1998.

\bibitem{HaTs98}
S.~Hanly and D.~Tse, ``Multiaccess fading channels. ii. delay-limited
  capacities,'' \emph{Information Theory, IEEE Transactions on}, vol.~44,
  no.~7, pp. 2816--2831, Nov 1998.

\bibitem{CaTaBi99}
G.~Caire, G.~Taricco, and E.~Biglieri, ``Optimum power control over fading
  channels,'' \emph{Information Theory, IEEE Transactions on}, vol.~45, no.~5,
  pp. 1468--1489, Jul 1999.

\bibitem{HwMaGaCi07}
C.-S. Hwang, M.~Malkin, A.~El~Gamal, and J.~M. Cioffi, ``Multiple-access
  channels with distributed channel state information,'' in \emph{ISIT}, june
  2007, pp. 1561 --1565.

\bibitem{GamalKim11}
A.~El~Gamal and Y.-H. Kim, \emph{Network Information Theory}.\hskip 1em plus
  0.5em minus 0.4em\relax Cambridge University Press, 2011.

\bibitem{DPD11}
Y.~Deshpande, S.~R.~B. Pillai, and B.~K. Dey, ``On the sum capacity of
  multiaccess block-fading channels with individual side informatioin,'' in
  \emph{IEEE Information Theory Workshop, Paraty}, 2011.

\bibitem{MiFrTs12}
P.~Minero, M.~Franceschetti, and D.~Tse, ``Random access: An
  information-theoretic perspective,'' \emph{Information Theory, IEEE
  Transactions on}, vol.~58, no.~2, pp. 909 --930, feb. 2012.

\bibitem{IPD12}
K.~Iyer, S.~R.~B. Pillai, and B.~K. Dey, ``Power controlled adaptive sum
  capacity in the presence of distributed csi,'' in \emph{IEEE ISITA}, 2012.

\bibitem{CovTho91}
T.~M. Cover and J.~A. Thomas, \emph{Elements of Information Theory}.\hskip 1em
  plus 0.5em minus 0.4em\relax Wiley, 1991.

\bibitem{SibiHanly10}
S.~Bhaskaran, S.~Hanly, N.~Badruddin, and J.~Evans, ``Maximizing the sum rate
  in symmetric networks of interfering links,'' \emph{Information Theory, IEEE
  Transactions on}, vol.~56, no.~9, pp. 4471--4487, Sept 2010.

\bibitem{SDP13}
S.~Sreekumar, B.~K. Dey, and S.~R.~B. Pillai, ``Adaptive sum-capacity in
  presence of distributed {CSI} for non-identical links,'' in
  \emph{International Symposium on Information Theory, ISIT, Istanbul}, 2013.

\bibitem{KnoppHumblet95}
R.~Knopp and P.~Humblet, ``Information capacity and power control in
  single-cell multiuser communications,'' in \emph{ICC '95 Seattle}, vol.~1,
  Jun. 1995, pp. 331 --335.

\end{thebibliography}

\nocite{}

\section{Appendix}

\begin{appendices}
\section{Derivation of the rate expressions in theorem \ref{thm:cont:capa} for continuous valued distributions}\label{disc2cont}
\begin{figure}[htbp]
\centering
\includegraphics[scale=0.5]{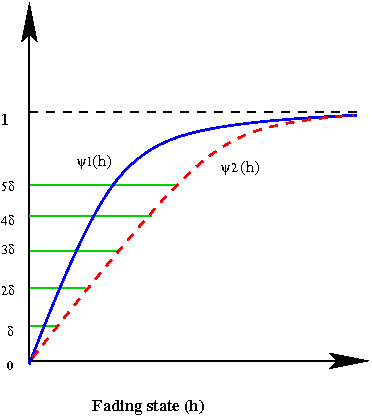}
\caption{Extension of the algorithm from discrete to continuous fading distributions}
\label{confi}
\end{figure}

We now show that the rate expressions given in theorem \ref{thm:cont:capa} for the continuous fading  
distribution can indeed be obtained by discretizing the CDF in the probability space and
 applying the rate allocation algorithm for the discrete fading state case given in theorem \ref{th:discrete}. 
Consider   fig.~\ref{confi}, which shows the CDF of the fading 
magnitudes $\psi_1(h_1)$ and $\psi_2(h_2)$ of user $1$ and user $2$ respectively. 
 Let $\delta$ be the interval between the uniform consecutive cuts as  shown in the figure.
Let $\psi_i^{-1}(.)$ denote the inverse CDF function defined by
\begin{align}
 \psi_i^{-1}(x)=h_i(x)=\sup\{g| \psi_i(g) < x \}
\end{align}
 
We apply the rate allocation given by theorem \ref{th:discrete} as follows.
For each horizontal cut $i$ shown in the figure, user $2$ selects rate $R_2(\psi_2\textsuperscript{-1}(i\delta))$ such that 
\begin{align}
 R_2(\psi_2\textsuperscript{-1}(i\delta)) + R_1(\psi_1\textsuperscript{-1}((i-1)\delta))&=\dfrac{1}{2}\log(1+(\psi_2\textsuperscript{-1}(i\delta))^2P_2+(\psi_1\textsuperscript{-1}((i-1)\delta))^2P_1) \label{cntfdeq1}
\end{align}
 In other words, user $2$ selects the rate $R_2(\psi_2\textsuperscript{-1}(i\delta))$ such that its sum with  $R_1(\psi_1\textsuperscript{-1}((i-1)\delta))$ achieves the sum rate constraint. 
In a similar manner, user $1$ selects its rate $R_1(\psi_1\textsuperscript{-1}((i-1)\delta))$ such that 
\begin{align}
 R_1(\psi_1\textsuperscript{-1}((i-1)\delta))+R_2(\psi_2\textsuperscript{-1}((i-1)\delta))&=\dfrac{1}{2}\log(1+P_2(\psi_2\textsuperscript{-1}((i-1)\delta))^2+P_1(\psi_1\textsuperscript{-1}((i-1)\delta))^2) \label{cntfdeq2}
\end{align}
This iterative assignment of rates to the users can be used to obtain a closed form expression for the rates
 in the limit $\delta$ tends to $0$ as we show below.

Subtracting \eqref{cntfdeq2} from  \eqref{cntfdeq1}, we get
\begin{align*}
&R_2(\psi_2\textsuperscript{-1}(i\delta))-R_2(\psi_2\textsuperscript{-1}((i-1)\delta))\\
&=\dfrac{1}{2}(\log(1+(P_2\psi_2\textsuperscript{-1}(i\delta))^2+P_1(\psi_1\textsuperscript{-1}((i-1)\delta))^2)-\dfrac{1}{2}\log(1+(P_2\psi_2\textsuperscript{-1}((i-1)\delta))^2+(P_1\psi_1\textsuperscript{-1}((i-1)\delta))^2)).
\end{align*}

Summing over i, we get
\begin{align}
 &R_2(\psi_2\textsuperscript{-1}(i\delta))-R_2(\psi_2\textsuperscript{-1}(0)) \notag \\
&= \sum_{j=1}^{i} \left(\dfrac{1}{2}(\log(1+P_2(\psi_2\textsuperscript{-1}(j\delta))^2+P_1(\psi_1\textsuperscript{-1}((j-1)\delta))^2) -\dfrac{1}{2}\log(1+P_2(\psi_2\textsuperscript{-1}((j-1)\delta))^2+P_1(\psi_1\textsuperscript{-1}((j-1)\delta))^2))\right) \label{cntfdeq3}
\end{align}

Approximating the difference term in the summation in \eqref{cntfdeq3} by partial derivatives, we get 
\begin{align*}
 &R_2(\psi_2\textsuperscript{-1}(i\delta))-R_2(\psi_2\textsuperscript{-1}(0))= \sum_{j=1}^{i} \dfrac{P_2\psi_2\textsuperscript{-1}((j-1)\delta)~d(\psi_2\textsuperscript{-1}((j-1)\delta))}{1+(P_2\psi_2\textsuperscript{-1}((j-1)\delta))^2+P_1(\psi_1\textsuperscript{-1}((j-1)\delta))^2} 
\end{align*}

where $d(.)$ denotes the differential. Now letting $\psi_2\textsuperscript{-1}((j-1)\delta)=y$ and $\psi_2\textsuperscript{-1}((i)\delta)=h$ and taking limit $\delta$ tends to $0$
, we get the expression for $R_2(h)$. In a similar manner, $R_1(h)$ is also obtained.

\section{Proof of Claim~\ref{claim:one}} \label{app:sec3c}
\allowdisplaybreaks{
\begin{align*}
 R_i(h_i)&=R_i(\mia)+\int_{\mia}^{h_i}\dfrac{yP_i}{1+y^2P_i +(\psi_{-i}^{-1}(\psi_i(y)))^2P_{-i}}\,\mathrm{d}y\\
& \leq R_i(\mia)+\int_{\mia}^{h_i}\dfrac{yP_i}{1+y^2P_i}\,\mathrm{d}y\\
&=R_i(\mia)+\int_{1+\mias P_i}^{1+h_i^2P_i}\frac{1}{2p} \mathrm{d}p\\
&=R_i(\mia)+\frac12 \log(1+h_i^2P_i)-\frac12 \log(1+\mias P_i) \\
&\leq \frac12 \log(1+\mias P_i) +\frac12 \log(1+h_i^2P_i)-\frac12 \log(1+\mias P_i) \\
&=\frac12 \log(1+h_i^2P_i)
\end{align*}
}

Now, let $R_{12}$ denote the sum-rate $R_1(h_1)+R_2(h_2)$, where the rates are
chosen as in \eqref{eq:rate:cont}. Under the transformation $\psi_2^{-1}\left(\psi_1(y)\right) = z$, we have
\begin{align*}
R_{12}	&=R_1(h_1(0))+R_2(h_2(0))+ \int\limits_{h_2(0)}^{\psi_2\textsuperscript{-1}(\psi_1(h_1))} 
	\dfrac{P_1\psi_1\textsuperscript{-1}(\psi_2(z))
	(\psi_1\textsuperscript{-1}(\psi_2(z))' }
	{1+z^2P_2+(\psi_1\textsuperscript{-1}(\psi_2(z)))^2P_1}\,dz 
	+ \int\limits_{h_2(0)}^{h_2} 
	\dfrac{yP_2}{1+y^2P_2+(\psi_1\textsuperscript{-1}(\psi_2(y)))^2P_1}\,dy. 
\end{align*}
Consider the case when $\psi_2\textsuperscript{-1}(\psi_1(h_1))<h_2$. Combining
terms of the two integrals above,
\begin{align*}
R_{12}=
 R_1(h_1(0))+R_2(h_2(0))& +\int\limits_{h_2(0)}^{\psi_2\textsuperscript{-1}(\psi_1(h_1))} 
	\dfrac{P_1\psi_1\textsuperscript{-1}(\psi_2(z))
	(\psi_1\textsuperscript{-1}(\psi_2(z)))'+zP_2 }
	{1+z^2P_2+(\psi_1\textsuperscript{-1}(\psi_2(z)))^2P_1}\,dz \notag \\
	 & +\int\limits_{\psi_2\textsuperscript{-1}(\psi_1(h_1))}^{h_2} \dfrac{yP_2}{1+y^2P_2+(\psi_1\textsuperscript{-1}(\psi_2(y)))^2P_1}\,dy 
\end{align*}
We now substitute $1+z^2P_2+(\psi_1\textsuperscript{-1}(\psi_2(z)))^2P_1=p$
in the first integral. We also upper bound the second integral
by replacing $z$ in the third term of the denominator by the lower
limit of the integral. This gives an upper bound since $(\psi_1\textsuperscript{-1}(\psi_2(y)))^2$ is a non-decreasing function of $z$.
By denoting $h^*= (\psi_2\textsuperscript{-1}(\psi_1(h_1)))^2P_2+h_1^2P_1+1$,
we then get
\begin{align*}
R_{12}&\leq R_1(h_1(0))+R_2(h_2(0))+ \int\limits_{1+h_2^2(0)P_2+h_1^2(0)P_1}^{h^*} \dfrac{1}{2p}\,dp 
 +\int\limits_{\psi_2\textsuperscript{-1}(\psi_1(h_1))}^{h_2} \dfrac{yP_2}{1+y^2P_2+h_1^2P_1}\,dy  \\
&\leq \frac 12 \log(1+h_1^2(0)P_1+h_2^2(0)P_2) + \int\limits_{1+h_2^2(0)P_2+h_1^2(0)P_1}^{h^*} \dfrac{1}{2p}\,dp 
 +\int\limits_{\psi_2\textsuperscript{-1}(\psi_1(h_1))}^{h_2} \dfrac{yP_2}{1+y^2P_2+h_1^2P_1}\,dy  \\
& = \frac 12 \log(1+h_1^2(0)P_1+h_2^2(0)P_2) + \left(\frac 12 \log(h^*) - \frac 12 \log(1+h_1^2(0)P_1+h_2^2(0)P_2)\right)  \notag \\
& \hspace*{49mm}+ \left(\frac 12 (\log(1+h_2^2P_2+h_1^2P_1) - \log(h^*)\right) \notag \\
& = \frac 12 \log(1+h_2^2P_2+h_1^2P_1)  \notag 
\end{align*}
For the case $\psi_2\textsuperscript{-1}(\psi_1(h_1)) \geq h_2$, the
proof follows in a similar fashion.

\section{Proof of Theorem~\ref{thm:cont:mult}}
\label{sec:multderive}
Let us first find an upper bound for the expected sum-rate of any achievable scheme.
\begin{align}
\sum_{i=1}^N \eE[R_i(H_i)] = 
        \sum_{i=1}^N\int_{0}^{\infty} & R_i(h_i) d\psi_i(h_i)
\end{align}
Using similar steps as in the discrete-state derivation in \eqref{eq:thpf3}, we get
by the sum-rate bound of the $N$ user MAC capacity region (see \cite{GamalKim11}),
\begin{align}
\sum_{i=1}^N \eE[R_i(H_i)] 
        \leq \int\limits_0^1 \frac 12 \log(1+\sum_{j=1}^{N}h_j^2(x)P_j)\,dx \notag,
\end{align}

thus obtaining an upper bound to the achievable sum-rate.

We will also show that this upper bound is in fact achieved by the rate-allocations
prescribed in Theorem~\ref{th:discretemult}. The rest of the
proof follows from Lemmas~\ref{lem:multwo} and \ref{lem:mulone} presented
below.

\begin{lemma} \label{lem:multwo}
For $x\in[0,1]$ and the rate allocation in \eqref{eq:adptcpmult},
$$
\sum_{i=1}^NR_i(h_i(x)) = \frac 12 \log(1+ \sum_{j=1}^Nh_j^2(x)P_j).
$$
\end{lemma}
\begin{IEEEproof}
From the rate allocation in  theorem \ref{th:discretemult}, it follows that
$$\sum_{i=1}^NR_i(h_i(0)) = \frac 12 \log(1+ \sum_{j=1}^Nh_j^2(0)P_j)$$
Also, for $x >0$, 
\begin{align}
\sum_{i=1}^NR_i(h_i(x)) &=\sum_{i=1}^NR_i(h_i(0))+\sum_{i=1}^N \int_{h_i(0)}^{h_i(x)} \frac{yP_idy}{1+\sum_{j=1}^{N}(\psi_{j}^{-1}(\psi_i(y)))^2P_{j}}\\
&=\frac 12 \log(1+ \sum_{j=1}^Nh_j^2(0)P_j)+\sum_{i=1}^N \int_{h_i(0)}^{h_i(x)} \frac{yP_idy}{1+\sum_{j=1}^{N}(\psi_{j}^{-1}(\psi_i(y)))^2P_{j}}.
\end{align}
Denoting, $\psi_1^{-1}(\psi_i(y))=z$ in the $i^{th}$ term in the summation and simplifying as discussed in lemma \ref{lem:two},
\begin{align*}
\sum_{i=1}^NR_i(h_i(x)) &=\frac 12 \log(1+ \sum_{j=1}^Nh_j^2(0)P_j)+\int\limits_{h_i(0)}^{h_1(x)} \frac{\sum_{k=1}^{N}(\psi_{k}^{-1}(\psi_1(y)))P_{k}}{1+\sum_{j=1}^{N}(\psi_{j}^{-1}(\psi_1(y)))^2P_{j}} ,\\
        &=\frac 12 \log(1+ \sum_{j=1}^Nh_j^2(0)P_j)+\int\limits_{1+ \sum_{j=1}^Nh_j^2(0)P_j}^{1+\sum_{k=1}^{N}h_k^2(x)P_k} \dfrac{1}{2p}\,dp \\
        &=\dfrac{1}{2}\log(1+\sum_{k=1}^{N}h_k^2(x)P_k).
\end{align*}
This completes the proof of the lemma.
\end{IEEEproof}

\begin{lemma} \label{lem:mulone}
The rate allocation given in \eqref{eq:adptcpmult} is outage-free.
\end{lemma}
\begin{IEEEproof}
We will show that $\forall (h_1,h_2,\cdots h_N) \in \bar H$ such that $h_i \geq h_{i}(0)$  and $ \forall S \subseteq \{1,2,\cdots,N\}$
\begin{align*}
 \sum_{i \in S}R_i(h_i)  &\leq \frac 12 \log(1+\sum_{i\in S} h_i^2 P_i ).
\end{align*}

Let $|S|$ denote the cardinality of the set $S$ and
 let $L$ be the vector formed by reading from left to right the indices of $\psi_i(h_i),~i \in S$ arranged in increasing order.
Let $L(l)$ denote the value of the $l^{th}$ component of $L$. Hence,  $\psi_{L(1)}(h_{L(1)}) \leq \psi_{L(2)}(h_{L(2)}) \leq \cdots \leq \psi_{L(|S|)}(h_{L(|S|)})$.

\begin{align}
 \sum_{i\in S}R_i(h_i) &=\sum_{i \in S}R_i(h_i(0))+\sum_{i\in S} \int_{h_i(0)}^{h_i} \frac{yP_idy}{1+\sum_{j=1}^{N}(\psi_{j}^{-1}(\psi_i(y)))^2P_{j}} \notag \\
& \leq \frac 12 \log(1+\sum_{i\in S} h_i^2(0) P_i )+\sum_{i=1}^{|S|} \int_{h_{L(i)}(0)}^{h_{L(i)}} \frac{yP_{L(i)}dy}{1+\sum_{j=1}^{N}(\psi_{j}^{-1}(\psi_{L(i)}(y)))^2P_{j}}\notag\\
&\leq \frac 12 \log(1+\sum_{i\in S} h_i^2(0) P_i )+\sum_{i=1}^{|S|} \int_{h_{L(i)}(0)}^{h_{L(i)}} \frac{yP_{L(i)}dy}{1+\sum_{j=1}^{|S|}(\psi_{L(j)}^{-1}(\psi_{L(i)}(y)))^2P_{L(j)}}\label{musrone}
\end{align}

As discussed in the two user case, substitute $\psi_1^{-1}(\psi_{L(i)}(y))=z$ in the $i^{th}$ term in \eqref{musrone} and simplify.
\begin{align}
 \sum_{i\in S}R_i(h_i) & \leq \frac 12 \log(1+\sum_{i\in S} h_i^2(0) P_i )+\sum_{i=1}^{|S|} \int_{h_{L(1)}(0)}^{\psi_{L(1)}^{-1}(\psi_{L(i)}(h_{L(i)}))} \frac{\psi_{L(i)}^{-1}(\psi_{L(1)}(y))P_{L(i)}dy}{1+\sum_{j=1}^{|S|}(\psi_{L(j)}^{-1}(\psi_{L(1)}(y)))^2P_{L(j)}}\notag \\
&=\frac 12 \log(1+\sum_{i\in S} h_i^2(0) P_i )+\sum_{k=1}^{|S|} \sum_{i=k}^{|S|}\int_{\psi_{L(1)}^{-1}(\psi_{L(k-1)}(h_{L(k-1)}))}^{\psi_{L(1)}^{-1}(\psi_{L(k)}(h_{L(k)}))} \frac{\psi_{L(i)}^{-1}(\psi_{L(1)}(y))P_{L(i)}dy}{1+\sum_{j=1}^{|S|}(\psi_{L(j)}^{-1}(\psi_{L(1)}(y)))^2P_{L(j)}}\notag \\
&=\frac 12 \log(1+\sum_{i\in S} h_i^2(0) P_i )+\sum_{k=1}^{|S|}\int_{\psi_{L(1)}^{-1}(\psi_{L(k-1)}(h_{L(k-1)}))}^{\psi_{L(1)}^{-1}(\psi_{L(k)}(h_{L(k)}))}\frac{\sum_{i=k}^{|S|}\psi_{L(i)}^{-1}(\psi_{L(1)}(y))P_{L(i)}dy}{1+\sum_{j=1}^{|S|}(\psi_{L(j)}^{-1}(\psi_{L(1)}(y)))^2P_{L(j)}} \label{ust2}
\end{align}

where $\psi_{L(1)}^{-1}(\psi_{L(0)}(h_{L(0)}))=h_{L(1)}(0)$. Also for $y \in \left[\psi_{L(1)}^{-1}(\psi_{L(k-1)}(h_{L(k-1)})),\psi_{L(1)}^{-1}(\psi_{L(k)}(h_{L(k)}))\right]$,

\begin{align}
 \psi_{L(1)}^{-1}(\psi_{L(j)}(h_{L(j)}))\leq \psi_{L(1)}^{-1}(\psi_{L(k-1)}(h_{L(k-1)})) \leq y,~\forall j \leq k-1. \label{ust1}
\end{align}

Hence in the  $k^{th}$ term in the outer summation, $\forall j \leq k-1$, substituting $\psi_{L(1)}^{-1}(\psi_{L(j)}(h_{L(j)}))$ for y in the $j^{th}$ term in the summation in the denominator of equation \eqref{ust2}, we get
\begin{align}
 \sum_{i\in S}R_i(h_i) & \leq \frac 12 \log(1+\sum_{i\in S} h_i^2(0) P_i )+\sum_{k=1}^{|S|}\int_{\psi_{L(1)}^{-1}(\psi_{L(k-1)}(h_{L(k-1)}))}^{\psi_{L(1)}^{-1}(\psi_{L(k)}(h_{L(k)}))}\frac{\sum_{i=k}^{|S|}\psi_{L(i)}^{-1}(\psi_{L(1)}(y))P_{L(i)}dy}{1+\sum_{j=1}^{k-1}h_{L(j)}^2P_{L(j)}+\sum_{j=k}^{|S|}(\psi_{L(j)}^{-1}(\psi_{L(1)}(y)))^2P_{L(j)}} \notag \\
&=\frac 12 \log(1+\sum_{i\in S} h_i^2(0) P_i )+\sum_{k=1}^{|S|}\int_{1+\sum_{j=1}^{k-1}h_{L(j)}^2P_{L(j)}+\sum_{j=k}^{|S|}(\psi_{L(j)}^{-1}(\psi_{L(k-1)}(h_{L(k-1)})))^2P_{L(j)}}^{1+\sum_{j=1}^{k-1}h_{L(j)}^2P_{L(j)}+\sum_{j=k}^{|S|}(\psi_{L(j)}^{-1}(\psi_{L(k)}(h_{L(k)})))^2P_{L(j)}}\frac{dp}{2p} \label{ust3} 
\end{align}

where the empty summation is defined to be 0, i.e. $\sum_{j=1}^0 h_{L(j)}^2P_{L(j)}=0$.

Simplifying \eqref{ust3},
\begin{align}
 \sum_{i\in S}R_i(h_i) &\leq \frac 12 \log(1+\sum_{i\in S} h_i^2(0) P_i )+\frac 12 \log(1+\sum_{i=1}^{|S|}h_{L(i)}^2P_{L(i)})-\frac 12 \log(1+\sum_{i\in S} h_i^2(0) P_i ) \notag\\
&=\frac 12 \log(1+\sum_{i\in S}h_i^2P_i)\notag
\end{align}

This proves the claim.

\end{IEEEproof}


\section*{ACKNOWLEDGMENTS}
This work was supported in part by Bharti Centre for Communication at IIT Bombay, grant  SB/S3/EECE/077/2013 from the Department of Science and Technology, Government of India, and a grant from the Information Technology Research Academy, Media Lab Asia, India. 

\end{appendices}

\end{document}